%% file: main.tex
\pdfoutput=1
\input{glyphtounicode}
\documentclass[sigconf, nonacm]{acmart}

\usepackage{amsmath}
\usepackage{amsthm}
\usepackage{array}
\usepackage{booktabs}
\usepackage{colortbl}
\usepackage{multirow}
\usepackage{graphicx}
\usepackage{xcolor}
\usepackage{algorithm}
\usepackage{algpseudocode}
\usepackage{enumitem}
\usepackage{balance}
\usepackage{bookmark}

\hypersetup{
  hidelinks,
  pdfauthor={Bowen Jiang, Xue Wang, Tao Yao},
  pdftitle={Fixed-Capacity Limits of Linearizable Mergeable Cardinality Sketches: Tight Bounds and CapPCSA}
}

\definecolor{rowhl}{HTML}{EEF3F8}
\definecolor{algsh}{HTML}{EEF3F8}
\newcommand{\rowhl}{\rowcolor{rowhl}}

\newcommand{\tblclean}{\setlength{\aboverulesep}{0pt}\setlength{\belowrulesep}{0pt}}

\DeclareRobustCommand{\texttt}[1]{\ifmmode\text{\upshape\ttfamily #1}\else{\upshape\ttfamily #1}\fi}

\newtheorem{theorem}{Theorem}
\newtheorem{lemma}{Lemma}
\newtheorem{proposition}{Proposition}
\newtheorem{corollary}{Corollary}
\newtheorem{remark}{Remark}

\newcommand{\mypara}[1]{\par\addvspace{1.2pt}\noindent\textbf{#1.}}
\newcommand{\cappcsa}{\textsc{CapPCSA}}

\input{tables/metrics_macros}

\begin{document}

\title{Fixed-Capacity Limits of Linearizable Mergeable Cardinality Sketches: Tight Bounds and CapPCSA}

\author{Bowen Jiang}
\affiliation{%
  \institution{Shanghai Jiao Tong University}
  \city{Shanghai}
  \country{China}
}
\email{bowenjiang@sjtu.edu.cn}

\author{Xue Wang}
\affiliation{%
  \institution{Independent Researcher}
  \country{United States}
}
\email{wxie91@gmail.com}

\author{Tao Yao}
\authornote{Corresponding author.}
\affiliation{%
  \institution{Shanghai Jiao Tong University}
  \city{Shanghai}
  \country{China}
}
\email{taoyao@sjtu.edu.cn}

\begin{abstract}
Strictly mergeable cardinality sketches evaluate \texttt{COUNT(DISTINCT)} across arbitrary aggregation trees with states bit-identical to sequential union. Variable-length or overflow-prone designs (e.g., Apache CPC and multi-cell Fishmonger) approach Pettie and Wang's linearizable Shannon-entropy floor of \Nm{c.pw21.3}. Yet fixed-width Parquet columns and bounded-slot frames require zero-overflow ($\delta{=}0$) fixed wire capacity, where prior fixed-size linearizable sketches suffer memory--variance products (MVPs) strictly above $3.00$ on 64-bit universes.

We close this gap with three contributions. First, we prove that deterministic fixed capacity raises the asymptotically tight linearizable limits to \textbf{\Nm{c.merge.5}} ($+\Nm{pct.above.pw21.1}\%$ above \Nm{c.pw21.3}) for inverse-Fisher MVP and \textbf{\Nm{c.rate.5}} ($+\Nm{pct.above.pwy21.1}\%$ above \Nm{c.pwy21.5}) for state-change rate, alongside an implication-DAG criterion and linearizable $\top$-collapse corollary. On matched 64-bit windows, a left-shifted integration window converges at $O((\ln\ln d)/d)$ and certifies non-vacuous bounds \textbf{\Nm{mw.lo.3}--\Nm{mw.hi.3}}. Second, we construct \cappcsa{}, a pooled-tail PCSA closure matching both limits asymptotically and certifying five 64-bit configurations with inverse-Fisher UB below $3.00$ ($+\Nm{mw.gap5.ex.lo}\%$--$+\Nm{mw.gap5.ex.hi}\%$ \textsc{Exact} / $+\Nm{mw.gap5.fa.lo}\%$--$+\Nm{mw.gap5.fa.hi}\%$ \textsc{Fast} above matched lower bounds), paired with zero-allocation $\delta{=}0$ codecs (\textsc{CapPCSA-Fast}, \Nm{ops.cap15.2k.fast.enc}--\Nm{ops.cap135.2k.fast.dec}\,$\mu$s at $g{=}2{,}048$ on M5; \textsc{RawEF}). Across 16 real-hash phases, \textsc{Fast} achieves \textbf{\Nm{ph.b16.fast135.mean2}--\Nm{ph.b16.fast15.mean2}} mean MVP (cutting \Nm{ph.cut.fast15.ell28}\%--\Nm{ph.cut.fast135.ell32}\% vs.\ ExaLogLog-$28$/$32$b). Dual-host and Parquet runs map the system trade-offs: \cappcsa{} leads fixed-wire catalog size and \texttt{FLBA} \texttt{NONE}/\texttt{LZ4} storage, ties ULL-8b and CPC on 16-phase \texttt{ZSTD} MVP, and trades higher working RAM and small-$n$ hashing cost for provable $\delta{=}0$. Third, we prove a continuous smoothed ExaLogLog relaxation diverges as $(\ln\Lambda + \ln\ln\Lambda)/I_0 + O(1)$ for $\Lambda{=}\log_2\log_2 U$.
\end{abstract}

\maketitle

\input{sections/01_intro}
\input{sections/02_model}
\input{sections/03_lower_bounds}
\input{sections/04_cappcsa}
\input{sections/05_ell_divergence}
\input{sections/06_evaluation}
\input{sections/07_related_conclusion}

\bibliographystyle{ACM-Reference-Format}
\bibliography{refs}

\clearpage
\appendix
\input{sections/appendix_merged}

\balance
\end{document}

%% file: sections/01_intro.tex
\section{Introduction}
\label{sec:intro}

Distributed analytical query engines, streaming telemetry pipelines, and catalog statistics managers rely on \emph{strictly mergeable} cardinality sketches to evaluate \texttt{COUNT(DISTINCT)} across arbitrary aggregation trees. A sketch is strictly mergeable if its state forms a join-semilattice under union: merging partial summaries over any partitioning or replication of a multiset yields the exact same bit-for-bit state as scanning the union sequentially. Space efficiency is governed by the \emph{memory--variance product} (MVP), $B \cdot \mathrm{RelVar}(\hat{\lambda})$, where $B$ is sketch size in bits and $\mathrm{RelVar}$ is relative mean-squared error.

Prior theory and practice leave a sharp gap between variable-length entropy sketches and deterministic fixed-size sketches. Pettie and Wang~\cite{pettie2021fish} proved that in the single-cell linearizable Dartboard model, Shannon entropy over Fisher information is asymptotically at least $H_0/I_0 \approx \Nm{c.pw21.3}$. Variable-length or overflow-prone sketches such as Apache DataSketches CPC~\cite{lang2017cpc,datasketches} and multi-cell Fishmonger~\cite{pettie2021fish} approach this entropy regime. Yet database engines (including Parquet \texttt{FIXED\_LEN\_BYTE\_ARRAY} columns, pinned shuffle slots, and fixed-schema catalog tables) often require a \emph{deterministic fixed wire capacity} ($|\mathcal{S}| \le 2^B$, zero overflow $\delta{=}0$) alongside exact mergeability. In that fixed-size regime, practitioners deploy independent register arrays such as HyperLogLog (HLL)~\cite{flajolet2007hll}, UltraLogLog (ULL)~\cite{ertl2024ull}, and ExaLogLog (ELL)~\cite{ertl2025ell}, or anchored-register closures~\cite{pettie2021nonmergeable} like our $k$-anchor Staircase-6b, which measure \Nm{s1m.curt4k.mvp2}--\Nm{s1m.hll.4k.mvp2} wire MVP at $n{=}10^6$. Neither the true fixed-capacity linearizable limit (already binding on 64-bit universes) nor a microsecond, zero-allocation $\delta{=}0$ closure breaking below $3.00$ was previously known.

\begin{figure*}[t]
  \centering
  \includegraphics[width=0.72\textwidth]{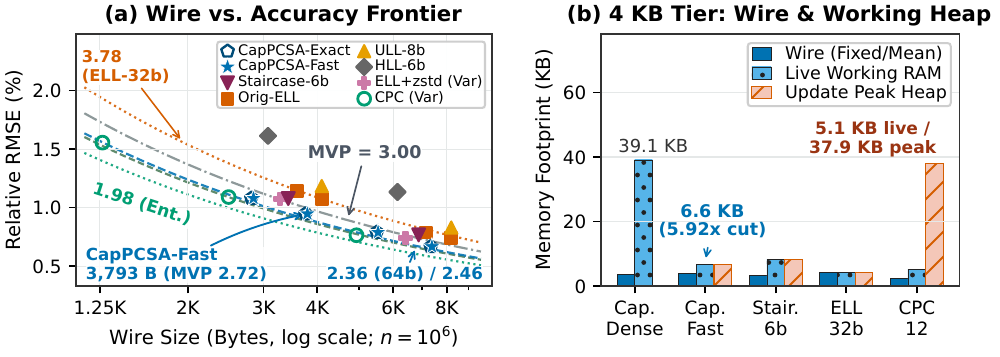}
  \Description{Two-panel plot showing the wire size versus relative RMSE frontier across cardinality sketches and a bar chart comparing wire size, live working RAM, and peak working heap at the 4 KB tier.}
  \caption{\textbf{\textsc{CapPCSA-Fast} measures \Nm{s1m.fast135.2k.mvp2} fixed-wire MVP at $n{=}10^6$ (\Nm{ph.b16.fast135.mean2} 16-phase mean; Arb-360 UB \Nm{fast.q135.2k.ub2}) and cuts RAM \Nm{cap.q135.2k.ramcut}$\times$.} \textbf{(a)} 4k-seed frontier ($n{=}10^6$, M5; \texttt{zstd}: 400-seed): four 2K/4K \textsc{Exact}/\textsc{Fast} ($+\Nm{fast.fig1.ovh.min}\%$--$\Nm{fast.fig1.ovh.max}\%$ bytes) points track window-level \Nm{c.merge.3}/\Nm{mw.hi.3} refs below Staircase-6b, Orig-ELL-32b, and ELL+\texttt{zstd}. \textbf{(b)} 4\,KB tier: sliding window cuts heap \Nm{cap.q135.2k.ramcut}$\times$ (\Nm{cap.q135.2k.dense} $\to$ \Nm{cap.q135.2k.ram}\,B; \Nm{cpc12.peak.ratio}$\times$ below CPC peak).}
  \label{fig:frontier}
\end{figure*}

\mypara{A Two-Level Taxonomy of Mergeable Cardinality Sketches}
Our results organize strictly mergeable sketches into two clean architectural levels (Figure~\ref{fig:frontier}):
\begin{enumerate}[leftmargin=*, topsep=2pt, itemsep=1.5pt]
  \item \textbf{Level 1: Variable-Length Entropy vs.\ Deterministic Fixed Capacity.} Variable-length or $\delta$-failure encodings (CPC~\cite{lang2017cpc,datasketches}, rollback Fishmonger~\cite[\S6]{pettie2021fish}) approach Pettie--Wang's linearizable entropy floor \Nm{c.pw21.3}~\cite[Thm.~6]{pettie2021fish} at the cost of variable framing or overflow failure. Deterministic fixed capacity ($\delta{=}0$) raises this limit by $+\Nm{pct.above.pw21.1}\%$ to \textbf{\Nm{c.merge.3}} asymptotically and \textbf{\Nm{mw.lo.3}--\Nm{mw.hi.3}} on matched 64-bit windows.
  \item \textbf{Level 2: Three Structural Realizations Inside Fixed Capacity.} Within $\delta{=}0$ fixed capacity, designs split by scale sharing:
  (2a)~\emph{Pooled-tail bitmap closures} (\cappcsa{}) share sparse tail budget across all columns to break below $3.00$;
  (2b)~\emph{Shared-anchor registers} (Curtain~\cite{pettie2021nonmergeable} evaluated at \Nm{curtain.3b.mvp} under our smoothed Fisher protocol, and $k$-anchor Staircase-6b at \Nm{st.b6.slot.lo}--\Nm{st.b6.worst.hi} theory / \Nm{ph.b16.curt.mean2} 16-phase empirical) face 1-Lipschitz limits (App.~C); and
  (2c)~\emph{Independent registers} (HLL~\cite{flajolet2007hll}, ULL~\cite{ertl2024ull}, ELL~\cite{ertl2025ell}) run sub-$\mu$s x86-64 merges in minimal RAM, but per-register rank bits force smoothed ELL divergence $(\ln\Lambda{+}\ln\ln\Lambda)/I_0{+}O(1)$ ($\Lambda{=}\log_2\allowbreak\log_2 U$).
\end{enumerate}

This taxonomy guides deployment. \textsc{Fast} ($\delta{=}0$, UB $<3.00$) targets fixed-schema Parquet \texttt{FLBA} (\texttt{NONE}/\texttt{LZ4}) catalogs and pinned RDMA shuffle buffers, whereas \textsc{RawEF} ($\delta{=}0$; UB \Nm{rawef.q135.4k.ub3}/\Nm{rawef.q125.16k.ub3} at $g{\ge}4\text{K}, q{\le}1.35$; $q{=}1.5,2$K UB \Nm{rawef.q15.2k.ub3} cuts dense OLAP CPU). Meanwhile, ELL and ULL lead x86-64 merges, RAM MVP, small-$n$ inserts, and raw \texttt{ZSTD} bytes (tying \cappcsa{} on 16-phase \texttt{ZSTD} MVP), while CPC leads variable-wire sparse streams.

\mypara{Contributions}
Unlike prior entropy bounds~\cite{pettie2021fish,pettie2021nonmergeable}, we replace Shannon entropy via stochastic domination ($Q_\lambda \le_{\mathrm{st}} P_\lambda$) charging $1$ bit above hit probability $1/2$, and integrate over a left-shifted window $[a{-}\tau, b]$ across three pillars:
\begin{itemize}[leftmargin=*, topsep=1pt, itemsep=1pt]
  \item \textbf{Pillar I (\S\ref{sec:model}--\S\ref{sec:lb}):} Theorem~\ref{thm:lb} proves $B/\!\inf_{[a,b]}\mathcal{J}(x) \ge \Nm{c.merge.5} - O((\ln\ln d)/d)$ ($+\Nm{pct.above.pw21.1}\%$ above \Nm{c.pw21.3}; matched 64-bit \textbf{\Nm{mw.lo.3}--\Nm{mw.hi.3}}, width factor \Nm{d64.mrg.arb.3}) and state-change rate \Nm{c.rate.5}. Proposition~\ref{prop:dag} characterizes linearizability via an implication DAG (Table~\ref{tab:applicability}), and Corollary~\ref{cor:top_collapse} covers linearizable $\top$-collapses.
  \item \textbf{Pillar II (\S\ref{sec:cappcsa}, \S\ref{sec:eval}):} Theorem~\ref{thm:cappcsa} constructs \cappcsa{} ($\delta{=}0$, asymptotically tight, five 64-bit configs with inverse-Fisher UB $<3.00$ at $+\Nm{mw.gap5.ex.lo}\%$--$+\Nm{mw.gap5.ex.hi}\%$ \textsc{Exact} / $+\Nm{mw.gap5.fa.lo}\%$--$+\Nm{mw.gap5.fa.hi}\%$ \textsc{Fast} above LBs; \texttt{q2p0} $+\Nm{mw.gap.q2.ex}\%$/$+\Nm{mw.gap.q2.fa}\%$). Propositions~\ref{prop:fast_codec}--\ref{prop:sliding} add zero-alloc \textsc{Fast} (\Nm{fast.lut.rans.min.kb}--\Nm{fast.lut.rans.max.kb}\,KB rANS L1 LUTs inside \Nm{fast.static.min.kb}--\Nm{fast.static.max.kb}\,KB static LUTs; \Nm{fast.spd.min}$\times$--\Nm{fast.spd.max}$\times$ M5 / \Nm{xeon.spd.min}$\times$--\Nm{xeon.spd.max}$\times$ Xeon speedup) and \textsc{RawEF}. This cuts working RAM \Nm{ramcut.min}$\times$--\Nm{ramcut.max}$\times$ and achieves \textbf{\Nm{ph.b16.fast135.mean2}--\Nm{ph.b16.fast15.mean2}} 16-phase MVP (\Nm{ph.cut.fast15.ell28}\%--\Nm{ph.cut.fast135.ell28}\% below ELL-28b, \Nm{ph.cut.fast15.ell32}\%--\Nm{ph.cut.fast135.ell32}\% below ELL-32b, \Nm{ph.cut.fast15.curt}\%--\Nm{ph.cut.fast135.curt}\% below Staircase-6b).
  \item \textbf{Pillar III (\S\ref{sec:ell}):} Theorem~\ref{thm:ell_div} proves smoothed ELL diverges as $F^*(U) = (\ln\Lambda + \ln\ln\Lambda)/I_0 + O(1)$ ($\Lambda{=}\log_2\log_2 U$, \Nm{ell.fstar.64} at 64-bit; Figure~\ref{fig:ell_div}).
\end{itemize}

%% file: sections/02_model.tex
\section{Model, Linearizability, and Scope}
\label{sec:model}

We formalize the single-cell Poissonized Dartboard model, prove an exact graph-theoretic criterion for linearizability, and audit classical and modern sketches under their native single-key arrival laws.

\subsection{Single-Cell Poissonized Dartboard Model}
Following Pettie and Wang~\cite{pettie2021fish}, let $\mathcal{C} = \{c_0, \dots, c_{N-1}\}$ be a finite cell partition of the Dartboard $[0,1]^2$ with strictly positive cell areas $p_0, \dots, p_{N-1} > 0$ summing to $1$. (For a finite-$K$ geometric grid of total active mass $1 - q^{-K} < 1$, the residual mass $q^{-K}$ sits on a permanently forced-occupied, zero-information dummy cell $c_\infty$ so $\sum_i p_i = 1$ holds identically.) Under the Poissonized random-oracle arrival law $P_\lambda$ at cardinality parameter $\lambda = e^x > 0$, raw cell hit indicators $Z_0, \dots, Z_{N-1}$ are mutually independent with
\begin{equation}
\label{eq:poisson_hits}
  Z_i \sim \mathrm{Bernoulli}\!\left(1 - e^{-p_i \lambda}\right), \qquad t_i(\lambda) := p_i \lambda = p_i e^x.
\end{equation}

For an extensive map $\Phi: \{0,1\}^N \to \{0,1\}^N$ ($X \subseteq \Phi(X)$), Pettie and Wang~\cite[Thm.~5]{pettie2021fish} proved that the strict union-homomorphism identity $\Phi(\Phi(X) \cup \Phi(W)) = \Phi(X \cup W)$ for all $X, W \in \{0,1\}^N$ holds if and only if $\Phi$ is a \emph{closure operator} (extensive, coordinatewise monotone, and idempotent). Equivalently, $\Phi$ is the minimal-enclosing-state map $X \mapsto \bigcap\{S \in \mathcal{S} : X \subseteq S\}$ of a family $\mathcal{S} \subseteq \{0,1\}^N$ containing $\mathbf{1}$ and closed under coordinatewise intersection, making $(\mathcal{S}, \vee)$ a finite join-semilattice with $S_1 \vee S_2 = \Phi(S_1 \cup S_2)$.

A closure operator $\Phi$ is \emph{linearizable} on generator support $G = \mathcal{C}$ if there exists a fixed permutation $\pi = (c_{\pi(0)}, \dots, c_{\pi(N-1)})$ of $\mathcal{C}$ and parameter-independent, prefix-monotone Boolean forced-occupation functions $\phi_k: \{0,1\}^k \to \{0,1\}$ such that $Y = \Phi(Z)$ satisfies
\begin{equation}
\label{eq:lin_def}
  Y_{\pi(k)} = Z_{\pi(k)} \vee \phi_k\!\left(Y_{\pi(0)}, \dots, Y_{\pi(k-1)}\right) \quad \forall k \in \{0,\dots,N-1\}.
\end{equation}
Let $h_{\pi(k)}(x) := \Pr_{P_{e^x}}(\phi_k(Y_{\pi(<k)}) = 0) \in [0,1]$ denote cell $c_{\pi(k)}$'s visibility (unforced) probability at log-cardinality $x = \ln\lambda$ ($k \in \{0,\dots,N{-}1\}$). Because $Z^{(x)}$ is stochastically non-decreasing in $x$ and every $\phi_k$ is monotone, each $h_i(x)$ is non-increasing in $x$. Define the atomic Shannon-entropy, Fisher-information, and state-change-rate kernels:
\begin{equation}
\label{eq:kernels}
  \dot{H}(t) := h_2(1{-}e^{-t}), \quad \dot{I}(t) := \frac{t^2}{e^t - 1}, \quad \dot{R}(t) := 2t e^{-t},
\end{equation}
whose log-integrals over $(0,\infty)$ are $H_0 = \int_0^\infty \dot{H}(u)\frac{du}{u} = \Nm{h0.6}\ldots$, $I_0 = \frac{\pi^2}{6} = \Nm{i0.6}\ldots$, and $\int_0^\infty \dot{R}(u)\frac{du}{u} = 2$. By the exact chain rules of~\cite[Lem.~13]{pettie2021fish} and~\cite[\S3.1]{pettie2021nonmergeable}, score cross-terms vanish under~\eqref{eq:lin_def}, yielding for every $x \in \mathbb{R}$:
\begin{align}
\label{eq:chain_rules}
  \mathcal{J}(x) &:= e^{2x} I_Y(e^x) = \sum_{i=0}^{N-1} \dot{I}(p_i e^x)\, h_i(x), \nonumber \\
  \mathcal{R}(x) &:= 2e^x \mathbb{E}_{P_{e^x}}[p_{\mathrm{chg}}(Y)] = \sum_{i=0}^{N-1} \dot{R}(p_i e^x)\, h_i(x),
\end{align}
where $p_{\mathrm{chg}}(Y) = \sum_{i : Y_i = 0} p_i$ is the probability that the next independent single-cell arrival changes state $Y$.

\noindent\textit{Notation disambiguation.} Throughout, we distinguish Theorem~\ref{thm:lb}'s window shift $\tau > 0$, \cappcsa{}'s floor level $L(X) \in \{0,\dots,K\}$ (\S\ref{sec:cappcsa}), and Theorem~\ref{thm:ell_div}'s universe scale $\Lambda := \log_2\log_2 U$ (\S\ref{sec:ell}); similarly $\Phi$ denotes a closure operator whereas $\Psi(u){:=}\dot{H}(u)/u$ (\S\ref{sec:lb}), and $d{=}b{-}a$ denotes log-window width whereas $d_{\mathrm{ell}}$ denotes ELL history bits.

\subsection{Acyclic Minimal-Implication Criterion and Applicability Audit}
To make the exact boundary of linearizability checkable for any concrete sketch, call $(A, i)$ with $A \subseteq G$ and $i \in G \setminus A$ a \emph{minimal closure implication} of $(\Phi, G)$ if $i \in \Phi(A)$ and $i \notin \Phi(A')$ for every proper subset $A' \subsetneq A$. Define the directed \emph{implication dependency graph} $D(\Phi, G) = (G, E)$ with edge $a \to i$ whenever there exists a minimal closure implication $(A, i)$ with $a \in A$.

\begin{proposition}[DAG Characterization of Linearizability]
\label{prop:dag}
Given a finite generator support $G$ and closure operator $\Phi: 2^G \to 2^G$, $\Phi$ is linearizable on $G$ if and only if $D(\Phi, G)$ is acyclic.
\end{proposition}

\noindent\textit{Relation to Horn knowledge bases.} Proposition~\ref{prop:dag} adapts Hammer and Kogan's classical acyclic Horn prime-implication theorem~\cite{hammer1995horn} to supported Dartboard closures, expressing closure directly as the topologically ordered occupancy recurrence~\eqref{eq:lin_def}.

\begin{proof}
($\Rightarrow$) Suppose $\Phi$ is linearizable in order $\pi = (g_0,\dots,g_{m-1})$ via~\eqref{eq:lin_def}. Let $(A, i)$ be a minimal closure implication and pick $a \in A$ with positions $k_a = \pi^{-1}(a)$ and $k_i = \pi^{-1}(i)$. Comparing $Z_1 = A \setminus \{a\}$ and $Z_2 = A$, both inputs agree on all coordinates before $k_a$, so induction on~\eqref{eq:lin_def} shows $\Phi(Z_1)$ and $\Phi(Z_2)$ agree on $\{g_j : j < k_a\}$. If $k_i < k_a$, then $i \in \Phi(A\setminus\{a\}) \iff i \in \Phi(A)$, contradicting minimality. Hence $k_a < k_i$ on every edge $a \to i$, so $\pi$ topologically orders $D(\Phi, G)$.

($\Leftarrow$) Conversely, let $\pi = (g_0,\dots,g_{m-1})$ be a topological order of acyclic $D(\Phi, G)$, define prefix-monotone maps $\phi_k(y_{<k}) := \mathbf{1}[g_k \in \Phi(\{g_j : j < k,\, y_j = 1\})]$, and set $\Psi(Z)_k = Z_k \vee \phi_k(\Psi(Z)_{<k})$. Induction using extensivity, monotonicity, and idempotence immediately yields $\Psi(Z) \subseteq \Phi(Z)$. For the reverse inclusion, suppose $\Phi(Z) \setminus \Psi(Z) \neq \emptyset$ and pick $g_k \in \Phi(Z) \setminus \Psi(Z)$ minimizing $k$. Since $Z \subseteq \Psi(Z)$, $g_k \notin Z$. Choose an inclusion-minimal $A \subseteq Z$ with $g_k \in \Phi(A)$. Then $(A, g_k)$ is a minimal implication, so every $a \in A$ precedes $g_k$ in $\pi$ and lies in $Z \subseteq \Psi(Z)$, forcing $\phi_k(\Psi(Z)_{<k}) = 1$ and $g_k \in \Psi(Z)$, a contradiction.
\end{proof}

Linearizability depends jointly on $(\mathcal{S}, \vee)$ and generator support $G$: on $B_2 = \{\bot, a, b, \top\}$ ($a \vee b = \top$), support $G = \{a, b\}$ has zero edges (acyclic), whereas adding $\top$ creates the 2-cycle $\top \to a \to \top$ (non-linearizable) and couples closed occupancies $Y_a, Y_b$. Table~\ref{tab:applicability} applies Proposition~\ref{prop:dag} to ten sketch families under native single-key arrivals.

\input{tables/tab_applicability}

\subsection{Deterministic Fixed Capacity and Finite-Universe Equivalence}
A sketch has \emph{deterministic fixed capacity $B > 0$ bits} ($\delta{=}0$) if $|\mathcal{S}| \le 2^B$, or $|\mathcal{S}_{\le U}| \le 2^B$ on $U$-key reachable states $\mathcal{S}_{\le U} := \{\Phi(Z) : \|Z\|_1 \le U\}$. For $B < U$, finite-universe reachability offers no escape from $|\mathcal{S}| \le 2^B$:

\begin{lemma}[Finite-Universe Support Eq{}uivalence]
\label{lem:finite_u}
For any closure $\Phi$ on $\{0,1\}^N$ and integer $U \ge 1$:
(1) If $B < U$ and $|\mathcal{S}_{\le U}| \le 2^B$, then $\mathcal{S} = \mathcal{S}_{\le U}$, so $|\mathcal{S}| \le 2^B$.
(2) If $B \ge U \ge 4e^{a+d}$, any closure with $\mathcal{J}(x) \ge \widetilde{I} > 0$ (resp.\ $\mathcal{R}(x) \ge \widetilde{R} > 0$) on $[a, a{+}d]$ satisfies $B/\widetilde{I} \ge 4e^d$ (resp.\ $B/\widetilde{R} \ge 2e^d$).
\end{lemma}

\begin{proof}
(1) Let $A \subseteq \mathcal{C}$ minimally generate $y \in \mathcal{S}$. All $2^{|A|}$ subsets $S \subseteq A$ have distinct closures (otherwise $a \in S_1 \setminus S_2$ gives $a \in \Phi(S_2) \subseteq \Phi(A\setminus\{a\}) \implies \Phi(A\setminus\{a\}) = y$). Thus $|A| \ge U$ would yield $2^U > 2^B$ distinct closures from subsets of any $U$-element subset $A_0 \subseteq A$, lying in $\mathcal{S}_{\le U}$; hence $|A| < U$ and $\mathcal{S} = \mathcal{S}_{\le U}$.
(2) Fisher data-processing ($\dot{I}(t) \le t$, $\sum_i p_i = 1$) gives $\widetilde{I} \le \mathcal{J}(a) \le e^a$, and $p_{\mathrm{chg}}(Y) \le 1$ gives $\widetilde{R} \le \mathcal{R}(a) \le 2e^a$; dividing $B \ge 4e^{a+d}$ proves both bounds.
\end{proof}

%% file: tables/tab_applicability.tex
\begin{table*}[t]
\centering
\caption{\textbf{\cappcsa{} satisfies single-cell $P_\lambda$, acyclic $D(\Phi,G)$, and $\delta{=}0$ fixed capacity.} Native arrival support $G$, Prop.~\ref{prop:dag} linearizability, and Thm.~\ref{thm:lb} coverage ($\mathrm{lgK}{:=}\log_2 K$ for CPC; 1-Lipschitz conds.\ in App.~C).}
\label{tab:applicability}
\footnotesize
\tblclean
\renewcommand{\arraystretch}{1.00}
\setlength{\tabcolsep}{3.5pt}
\begin{tabular}{@{} >{\raggedright\arraybackslash}p{2.30cm} >{\raggedright\arraybackslash}p{3.45cm} c c >{\raggedright\arraybackslash}p{2.75cm} >{\raggedright\arraybackslash}p{3.10cm} @{}}
\toprule
\textbf{Sketch Family} & \textbf{Single-Key Support $G$ \& State} & \textbf{Single-Cell $P_\lambda$?} & \textbf{$D(\Phi,G)$ DAG?} & \textbf{Capacity \& Merge Contract} & \textbf{Theorem~\ref{thm:lb} Coverage} \\
\midrule
\textbf{HLL}~\cite{flajolet2007hll} (64-bit, 6b, $m{=}2^p$) &
$(r,\rho)\in[m]{\times}[K_{\mathrm{HLL}}(p)]$, $K_{\mathrm{HLL}}(p){=}\min(63,65{-}p)$ &
Yes & Yes ($\downarrow\rho$) & Fixed $B{=}6m$; exact $\max$ ($\delta{=}0$) &
\textbf{Yes} (exact member on finite $p$-dependent $G$) \\
\addlinespace[1.5pt]
\textbf{ULL}~\cite{ertl2024ull} (64-bit, 8b, $p\ge 3$) &
$(r,\rho)\in[m]{\times}[65{-}p]$ (\cite[Alg.~3]{ertl2024ull} requires $p{\ge}3$; 6b max + 2b hist) &
Yes & Yes ($\downarrow\rho$) & Fixed $B{=}8m$; exact $\max{+}$shift-OR &
\textbf{Yes} (exact member for $p{\ge}3$) \\
\addlinespace[1.5pt]
\textbf{Orig-ELL}~\cite{ertl2025ell} (64-bit, $p{\ge}2, t, d$) &
$(r,k)\in[m]{\times}[2^t(65{-}p{-}t)]$ ($4(63{-}p)$ at $t{=}2$); $q_{\mathrm{bits}}{+}d$ bits &
Yes & Yes ($\downarrow k$) & Fixed $B{=}m(q_{\mathrm{bits}}{+}d)$; \cite[Alg.~5]{ertl2025ell} &
\textbf{Yes} (exact member; Thm.~\ref{thm:ell_div} proves continuous divergence) \\
\addlinespace[1.5pt]
\textbf{PCSA}~\cite{flajolet1985pcsa} ($g{\times}K$ bitmap) &
Explicit $(r,\ell)\in[g]{\times}[K]$ (residual $2^{-K}$ on dummy $c_\infty$) &
Yes & Yes (0 edges) & Fixed $B{=}gK$; bitwise OR ($\delta{=}0$) &
\textbf{Yes} (pays $O(\log U)$ bits/col) \\
\addlinespace[1.5pt]
\rowhl \textbf{\cappcsa{}} (this work) &
Explicit $(r,\ell)\in[g]{\times}[K]$, $p_\ell{=}(1{-}1/q)\,g^{-1} q^{-(\ell-1)}$ (+ dummy $c_\infty$ mass $q^{-K}$) &
Yes & Yes ($\downarrow\ell, s{\ge}1$) & Fixed wire bytes (\textbf{Exact}, \textbf{Fast}, \textbf{RawEF}; $\delta{=}0$) &
\textbf{Yes} (exact member \& \textbf{asymptotically tight}) \\
\addlinespace[1.5pt]
\textbf{Staircase-6b (ours)} &
$k$-anchor $A{+}$offsets$+$hist.\ (App.~C); Curtain~\cite{pettie2021nonmergeable} 3b/col &
Yes & Yes (App.~C) & Fixed array ($\delta{=}0$); exact OR-max &
\textbf{Yes} (\Nm{ph.b16.curt.mean2} 16-ph; theory \Nm{st.b6.slot.lo}--\Nm{st.b6.worst.hi}; 3b Curtain~\cite{pettie2021nonmergeable} eval.\ \Nm{curtain.3b.mvp}, fam.\ \Nm{curtain.int.mvp}--\Nm{curtain.3b.mvp}) \\
\midrule
\textbf{DLL / AVLL}~\cite{bushnell2026dll,bushnell2026avll} &
Shared floor + 4b / base-56 relative offsets &
Order-dep. & \textbf{No} & Fixed slot; merge overflow (non-exact) &
\textbf{No} (async tier promotion breaks union) \\
\addlinespace[1.5pt]
\textbf{CPC}~\cite{lang2017cpc,datasketches} (abstract vs.\ C++ v5.2.0) &
Abstract: $(r,c)\in[K]{\times}[64]$. Native C++: MurmurHash3, $\mathrm{lgK}\in[4,26]$ &
Abstract: Yes & Abstract: Yes & \textbf{Variable wire} (emp.\ 99.9\% table for $\mathrm{lgK}{\le}19$; uncomp.\ ${\ge}8K$\,B) &
\textbf{No} (var-wire tracks Pettie--Wang~\cite{pettie2021fish} \Nm{c.pw21.5}; uncomp.\ $\to$ PCSA) \\
\addlinespace[1.5pt]
\textbf{Fishmonger}~\cite{pettie2021fish} (STOC'21 \S6) &
Native~\cite[\S6]{pettie2021fish}: multi-cell bitmap OR; rollback or entropy wire &
\textbf{No} (coupled multi-cell) & Single-cell ideal: Yes & \textbf{Variable wire or $\delta{>}0$ rollback failure} &
\textbf{No} (fails single-cell \& $\delta{=}0$; ideal $\top$-collapse $\to$ Cor.~\ref{cor:top_collapse}) \\
\addlinespace[1.5pt]
\textbf{HyperBitBit64}~\cite{janson2025hypertwobits} &
Two 64-bit masks + mutable threshold $U$ &
Order-dep. & \textbf{No} & Fixed 134\,b; \textbf{approximate merge} &
\textbf{No} (not a closure operator) \\
\bottomrule
\end{tabular}
\end{table*}

%% file: sections/03_lower_bounds.tex
\section{Pillar I: Tight Fixed-Capacity Lower Bounds}
\label{sec:lb}

We prove that enforcing a deterministic fixed state-space capacity $|\mathcal{S}| \le 2^B$ raises the optimal inverse-Fisher constant from Pettie and Wang's \Nm{c.pw21.5} to \Nm{c.merge.6}, and the asymptotic state-change rate constant from \Nm{c.pwy21.5} to \Nm{c.rate.6}. The core proof replaces raw Shannon entropy under $P_\lambda$ with a stochastically dominated coupling $Q_\lambda \le_{\mathrm{st}} P_\lambda$ that charges one full bit to every visible cell whose occupancy exceeds $1/2$, then integrates over a \emph{left-shifted log-cardinality window} $[a{-}\tau, b]$ to eliminate coarse/fine boundary loss. On a 64-bit universe ($d = \ln(2^{64}/4) \approx \Nm{d64.val}$), this yields non-vacuous config-matched bounds \textbf{\Nm{mw.lo.3}--\Nm{mw.hi.3}} ($+\Nm{mw.pct.lo}\%$--$+\Nm{mw.pct.hi}\%$ above \Nm{c.pw21.5}; width factor \Nm{d64.mrg.arb.3} on $[1,2^{62}]$) and converges at $O((\ln\ln d)/d)$.

\subsection{Monotone Bit-Cost Kernel and Variational Minimum}
Define the \emph{monotone worst-case bit-cost kernel} $\widetilde{H}(t)$ and its piecewise cumulative log-integral $G(t)$:
\begin{equation}
\label{eq:htilde}
  \widetilde{H}(t) := h_2\!\left(\min\!\left(1 - e^{-t}, \tfrac{1}{2}\right)\right) =
  \begin{cases}
    \dot{H}(t), & 0 < t \le \ln 2, \\
    1, & t > \ln 2,
  \end{cases}
\end{equation}
\begin{equation}
\label{eq:G_def}
  G(t) := \int_0^t \widetilde{H}(u)\,\frac{du}{u} =
  \begin{cases}
    \int_0^t \dot{H}(u)\,\frac{du}{u}, & 0 < t \le \ln 2, \\
    H_{\mathrm{upper}} + \ln(t / \ln 2), & t > \ln 2,
  \end{cases}
\end{equation}
where $G(\ln 2) = H_{\mathrm{upper}} := \int_0^{\ln 2} \dot{H}(u)\frac{du}{u} = \Nm{hupper.6}\ldots$ (the logarithmic branch holds strictly for $t > \ln 2$; extending it below $\ln 2$ creates spurious roots at \Nm{root.spur.merge} and \Nm{root.spur.rate}). For kernel $K \in \{\dot{I}, \dot{R}\}$, set $J_K(t) := \int_0^t K(u)\frac{du}{u}$, giving $J_{\dot{I}}(t) = \frac{\pi^2}{6} - \mathrm{Li}_2(e^{-t}) + t\ln(1-e^{-t})$ with $J_{\dot{I}}(\infty) = I_0 = \frac{\pi^2}{6}$, and $J_{\dot{R}}(t) = 2(1 - e^{-t})$ with $J_{\dot{R}}(\infty) = 2$.

\begin{lemma}[Uniq{}ue Global Variational Minimum]
\label{lem:var_min}
For both $K = \dot{I}$ and $K = \dot{R}$, the ratio $F_K(t) := G(t)/J_K(t)$ is strictly decreasing on $(0, \ln 2]$ and attains a unique global minimum on $(0,\infty)$ at a unique root $t_* \in (\ln 2, \infty)$:
\begin{enumerate}[leftmargin=*, topsep=1.5pt, itemsep=1.5pt]
  \item For $K = \dot{I}$, $t_* \approx \Nm{t.merge.6}$ uniquely solves $G(t)\dot{I}(t) = J_{\dot{I}}(t)$ on $(\ln 2, \infty)$, yielding
  \begin{equation}
    C_{\mathrm{fixed}}^{\mathrm{merge}} := \inf_{t > 0} \frac{G(t)}{J_{\dot{I}}(t)} = \frac{1}{\dot{I}(t_*)} = \frac{e^{t_*} - 1}{t_*^2} = \mathbf{\Nm{c.merge.6}}
  \end{equation}
  ($+\Nm{pct.above.pw21.1}\%$ above Pettie--Wang's $H_0/I_0 = \Nm{c.pw21.5}$).
  \item For $K = \dot{R}$, $t_*^{\mathrm{M}} \approx \Nm{t.rate.6}$ uniquely solves $G(t)\dot{R}(t) = J_{\dot{R}}(t)$ on $(\ln 2, \infty)$, yielding
  \begin{equation}
    C_{\mathrm{fixed}}^{\mathrm{rate}} := \inf_{t > 0} \frac{G(t)}{J_{\dot{R}}(t)} = \frac{1}{\dot{R}(t_*^{\mathrm{M}})} = \frac{e^{t_*^{\mathrm{M}}}}{2 t_*^{\mathrm{M}}} = \mathbf{\Nm{c.rate.6}}
  \end{equation}
  ($+\Nm{pct.above.pwy21.1}\%$ above Pettie--Wang--Yin's $H_0/2 = \Nm{c.pwy21.5}$~\cite{pettie2021nonmergeable}).
\end{enumerate}
\end{lemma}

\begin{proof}
Write $r_K(t) := \widetilde{H}(t)/K(t)$ and $\psi_K(t) := r_K(t)J_K(t) - G(t)$, so $F_K'(t) = \frac{K(t)}{t J_K(t)^2}\psi_K(t)$. On $(0,\ln 2]$, expanding $(1-e^{-t})\ln(1-e^{-t}) = -e^{-t}\sum_{n=1}^\infty \frac{1-e^{-nt}}{n(n+1)}$ yields
\[
  r_{\dot{R}}(t) = \frac{\dot{H}(t)}{\dot{R}(t)} = \frac{1}{2\ln 2}\!\left[1 + \sum_{n=1}^{\infty} \frac{1 - e^{-nt}}{n(n+1)\,t}\right],
\]
which decreases strictly term-by-term. Since $\dot{R}(t)/\dot{I}(t) = 2(1-e^{-t})/t$ also strictly decreases, $r_{\dot{I}}(t) = r_{\dot{R}}(t)(\dot{R}(t)/\dot{I}(t))$ strictly decreases on $(0,\ln 2]$. Thus $G(t) = \int_0^t r_K(u)K(u)\frac{du}{u} > r_K(t)J_K(t)$, so $\psi_K(t) < 0$ and $F_K'(t) < 0$ on $(0,\ln 2]$. On $(\ln 2, \infty)$, $\widetilde{H}(t)=1$ and $r_K(t)=1/K(t)$ is unimodal with turning point $t_{\mathrm{turn}} \in (\ln 2, \infty)$ (\Nm{t.turn.i} for $\dot{I}$; $1$ for $\dot{R}$). Since $\psi_K'(t) = r_K'(t)J_K(t)$, $\psi_K < 0$ on $(\ln 2, t_{\mathrm{turn}}]$, strictly increases on $(t_{\mathrm{turn}}, \infty)$, and diverges to $+\infty$, yielding a unique root $t_* > t_{\mathrm{turn}}$ and global minimum $F_K(t_*) = 1/K(t_*)$ (Figure~\ref{fig:theory_gap}(a)).
\end{proof}

\subsection{Shifted-Window Finite-\texorpdfstring{$d$}{d} Bounds and Asymptotic Tightness}

\begin{theorem}[Fixed-Capacity Linearizable Lower Bounds]
\label{thm:lb}
Fix any log-cardinality window $[a,b]$ of width $d := b - a > 0$, any linearizable Dartboard sketch $Y = \Phi(Z)$ with arbitrary positive cell weights $p_0,\dots,p_{N-1} > 0$ ($\sum_i p_i = 1$), and deterministic fixed-size space budget $B > 0$ bits (either $|\mathcal{S}| \le 2^B$, or $|\mathcal{S}_{\le U}| \le 2^B$ with $U \ge 4e^b$). Let $(K, C, J(\infty))$ be either:
\begin{itemize}[leftmargin=*, topsep=1pt, itemsep=1pt]
  \item \textbf{(a) Mergeable Fisher:} $K = \dot{I}$, $C = C_{\mathrm{fixed}}^{\mathrm{merge}} = \Nm{c.merge.6}$, $J(\infty) = \pi^2/6$;
  \item \textbf{(b) State-Change Rate (Write Churn):} $K = \dot{R}$, $C = C_{\mathrm{fixed}}^{\mathrm{rate}} = \Nm{c.rate.6}$, $J(\infty) = 2$ ($\mathcal{R}(x) = 2\lambda\mathbb{E}[p_{\mathrm{chg}}(Y)]$ is twice the instantaneous log-cardinality write rate, so $\frac{1}{2}\int_x^{x+\ln 2}\mathcal{R}(u)\,du$ is expected state writes per cardinality doubling; the Jensen-relaxed proxy of~\cite{pettie2021nonmergeable}).
\end{itemize}
If $\sum_{i=0}^{N-1} K(p_i e^x) h_i(x) \ge \widetilde{K} > 0$ for all $x \in [a,b]$, then for every shift $\tau > 0$, defining the universal shifted-interval infimum
\begin{equation}
\label{eq:ck_def}
  c_K(\tau) := \inf_{0 < \alpha < t < \infty} \frac{G(t) - G(\alpha e^{-\tau})}{J_K(t) - J_K(\alpha)},
\end{equation}
we have
\begin{equation}
\label{eq:finite_d_lb}
  \frac{B}{\widetilde{K}} \;\ge\; \frac{d}{d + \tau}\, c_K(\tau)
  \quad\text{and}\quad
  \frac{B}{\widetilde{K}} \;\ge\; \frac{d - \tau}{d}\, c_K(\tau) \;\;\text{(if $\tau < d$)}.
\end{equation}
Moreover, for every $\tau > 0$ satisfying $\tau \ge C(J(\infty) - J_K(\alpha_0))$ and $G(\ln 2) \le C J_K(\alpha_0)$ at $\alpha_0 := e^\tau \ln 2$, letting $u_- \in (0,\ln 2)$ solve $r_K(u_-) = C$ and $\alpha_1(\tau) \in (0,\alpha_0]$ be the first positive root of $G(\alpha e^{-\tau}) = C J_K(\alpha)$,
\begin{equation}
\label{eq:ck_elem}
  c_K(\tau) \;\ge\; c_K^{\mathrm{elem}}(\tau) := C\!\left(1 - \frac{G(\alpha_1(\tau) e^{-\tau})}{G(u_-)}\right) = C - \exp\!\bigl(-\Omega(e^\tau)\bigr).
\end{equation}
Choosing $\tau(d) = \ln\ln d + O(1)$ yields $\frac{B}{\widetilde{K}} \ge C - O\!\left(\frac{\ln\ln d}{d}\right)$ uniformly over all positive weight vectors $(p_i)$ and all $d \ge 3$.
\end{theorem}

\begin{proof}
By Lemma~\ref{lem:finite_u}, if $|\mathcal{S}_{\le U}| \le 2^B$ and $B \ge U$, then $B/\widetilde{K} \ge 2e^d$ ($4e^d$ for $\dot{I}$, $2e^d$ for $\dot{R}$) already exceeds~\eqref{eq:finite_d_lb}; otherwise $|\mathcal{S}| \le 2^B$.

\emph{Step 1 (Sub-dominated coupling $Q_\lambda \le_{\mathrm{st}} P_\lambda$).} For $\lambda = e^x$, let $Q_\lambda$ draw independent cell indicators $Z_i^{(Q)} \sim \mathrm{Bernoulli}(q_i(\lambda))$ with $q_i(\lambda) := \min(1 - e^{-p_i\lambda}, 1/2) \le 1 - e^{-p_i\lambda}$. Coupling $Z^{(Q)}$ and $Z^{(P)}$ via shared i.i.d.\ $U_i \sim \mathrm{Unif}[0,1]$ ensures $Z^{(Q)} \le Z^{(P)}$ coordinatewise. Since each $\phi_k$ in~\eqref{eq:lin_def} is monotone, the unforced event $\{\phi_k(Y_{\pi(<k)}) = 0\}$ is downward-closed, giving $\Pr_{Q_\lambda}(\phi_k(Y_{\pi(<k)}) = 0) \ge h_{\pi(k)}(x)$ for every $k$ and $x \in \mathbb{R}$.

\emph{Step 2 (Gibbs inequality + entropy chain rule).} Because $|\mathcal{S}| \le 2^B$ and $Z_{\pi(k)}^{(Q)}$ is independent of $Y_{\pi(<k)}^{(Q)}$ with entropy $h_2(q_{\pi(k)}(e^x)) = \widetilde{H}(p_{\pi(k)} e^x)$, every $x \in \mathbb{R}$ satisfies
\begin{equation}
\label{eq:gibbs_chain}
  B \;\ge\; \log_2 |\mathcal{S}| \;\ge\; H_{Q_\lambda}\!\left(\Phi(Z^{(Q)})\right) \;\ge\; \sum_{i=0}^{N-1} \widetilde{H}(p_i e^x)\, h_i(x).
\end{equation}

\emph{Step 3 (Left-shifted layer-cake integration).} Fix any shift $\tau > 0$. Because~\eqref{eq:gibbs_chain} holds for \emph{all} $x \in \mathbb{R}$, integrate~\eqref{eq:gibbs_chain} over $[a{-}\tau, b]$ (width $d{+}\tau$) while integrating $\sum_i K(p_i e^x) h_i(x) \ge \widetilde{K}$ over $[a, b]$ (width $d$). For each cell $i$, write $h_i(x) = \int_0^1 \mathbf{1}[h_i(x) \ge s]\,ds$ (non-increasing in $x$). For any level $s \in (0,1]$ whose superlevel set $\{x \in [a,b] : h_i(x) \ge s\} = [a, \xi_s]$ has positive length ($\xi_s \in (a,b]$), monotonicity of $h_i$ guarantees $h_i(x) \ge s$ on the entire extended interval $[a{-}\tau, \xi_s]$ as well. Substituting $u = p_i e^x$, $\alpha = p_i e^a$, and $t = p_i e^{\xi_s} > \alpha$, definition~\eqref{eq:ck_def} immediately gives
\begin{align*}
  \int_{a-\tau}^{\xi_s} \widetilde{H}(p_i e^x)\,dx
  &= G(t) - G(\alpha e^{-\tau}) \\
  &\ge c_K(\tau)\bigl(J_K(t) - J_K(\alpha)\bigr)
   = c_K(\tau)\!\int_a^{\xi_s} K(p_i e^x)\,dx.
\end{align*}
Integrating over $s \in (0,1]$ via Tonelli's theorem and summing over $i$ yields $(d{+}\tau)B \ge c_K(\tau)\,d\,\widetilde{K}$, proving the first bound in~\eqref{eq:finite_d_lb}; integrating~\eqref{eq:gibbs_chain} on $[a,b]$ and $K$ on $[a{+}\tau, b]$ gives $\frac{d-\tau}{d}c_K(\tau)$ verbatim.

\emph{Step 4 (Elementary 3-case bound and $O((\ln\ln d)/d)$ rate).} Fix $\tau > 0$ satisfying the two scalar conditions at $\alpha_0 = e^\tau \ln 2$. For any $0 < \alpha < t$:
\begin{itemize}[leftmargin=*, topsep=1pt, itemsep=1pt]
  \item \emph{Case 1 ($\alpha \ge \alpha_0$):} Here $\alpha e^{-\tau} \ge \ln 2$, so $G(t) - G(\alpha e^{-\tau}) = \tau + \ln(t/\alpha) \ge \tau$, while $J_K(t) - J_K(\alpha) \le J(\infty) - J_K(\alpha_0) \le \tau/C$, giving ratio $\ge C$.
  \item \emph{Case 2 ($\alpha_1(\tau) \le \alpha < \alpha_0$):} Write $\Delta(\alpha) := G(\alpha e^{-\tau}) - C J_K(\alpha)$ on $(0,\alpha_0]$ ($\Delta(0^+){=}0$, $\Delta(\alpha_0)\le 0$). Note $\Psi(u):=\dot{H}(u)/u = \frac{e^{-u}}{\ln 2}(1 + \sum_{n=1}^\infty \frac{1}{n(n+1)}\int_0^n e^{-tu}dt)$ is a Laplace transform of a positive measure on $[1,\infty)$ that is not a single point mass, so Cauchy--Schwarz gives $(\ln \Psi)'' > 0$, i.e., $u E_{\dot{H}}'(u) + 1 - E_{\dot{H}}(u) > 0$ for $E_f(x) := x f'(x)/f(x)$. Meanwhile both $-\alpha E_{\dot{R}}'(\alpha) = \alpha \ge 1 - E_{\dot{R}}(\alpha)$ and $-\alpha E_{\dot{I}}'(\alpha) > 1 - E_{\dot{I}}(\alpha)$ (since $-\alpha E_{\dot{I}}'-(1{-}E_{\dot{I}}) = 4e^{-\alpha}(\sinh^2(\alpha/2)-(\alpha/2)^2)/(1{-}e^{-\alpha})^2 > 0$, App.~A) hold for all $\alpha > 0$. Thus at any zero of $g_\tau(\alpha) := E_{\dot{H}}(\alpha e^{-\tau}) - E_K(\alpha)$, $\alpha g_\tau'(\alpha) > 0$, so $g_\tau$ crosses $0$ at most once ($-\!\to\!+$); hence $\Delta'(\alpha) = \frac{K(\alpha)}{\alpha}(\dot{H}(\alpha e^{-\tau})/K(\alpha) - C)$ has at most two sign changes ($+\!\to\!-\!\to\!+$). Because $\Delta > 0$ near $0^+$ and $\Delta(\alpha_0) \le 0$, $\Delta(\alpha) \le 0$ holds throughout $[\alpha_1(\tau), \alpha_0]$ for every admissible $\tau > 0$. Combined with Lemma~\ref{lem:var_min} ($G(t) \ge C J_K(t)$), $G(t) - G(\alpha e^{-\tau}) \ge C(J_K(t) - J_K(\alpha))$.
  \item \emph{Case 3 ($0 < \alpha < \alpha_1(\tau)$):} If $t \le u_-$, $r_K(u) \ge C$ on $(\alpha, t]$ gives ratio $\ge C$. If $c_K^{\mathrm{elem}}(\tau) \le 0$, $c_K(\tau) \ge 0$ holds; else ($c_K^{\mathrm{elem}}(\tau) > 0$, $t > u_-$) definition~\eqref{eq:ck_elem} enforces $G(\alpha_1(\tau) e^{-\tau}) < G(u_-)$, hence $\alpha_1(\tau) e^{-\tau} < u_-$, so $G(t) \ge \max(G(u_-), C J_K(t))$ yields $G(t) - G(\alpha e^{-\tau}) \ge G(t)[1 - G(\alpha_1(\tau)e^{-\tau})/G(u_-)] \ge c_K^{\mathrm{elem}}(\tau)(J_K(t) - J_K(\alpha))$.
\end{itemize}
As $\alpha \to 0^+$ with $z{=}\alpha e^{-\tau}$, $G(z) = \frac{z}{\ln 2}(2 - \ln z) + O(z^2|\ln z|)$ (App.~A) while $C J_K(\alpha) = c_0 \alpha + O(\alpha^2)$ ($c_0 = C$ for $\dot{I}$, $2C$ for $\dot{R}$), forcing $\alpha_1(\tau) = \exp(-(c_0 \ln 2)e^\tau + \tau + 2 + o(1))$ and $G(\alpha_1(\tau)e^{-\tau}) = \exp(-\Omega(e^\tau))$. Setting $\tau(d) = \max\{\tau_0, \ln(\frac{2}{c_0\ln 2}\ln d)\}$ balances $\tau/d = O((\ln\ln d)/d)$ against $O((\ln d)/d^2)$ for all $d \ge 3$; here $\tau_0{=}1.80$ ($\dot{I}$) and $1.40$ ($\dot{R}$) maximize the 64-bit elementary bound~\eqref{eq:ck_elem} (while smaller shifts $\tau{=}1.20 / 0.85$ already satisfy admissibility for Arb-360~\eqref{eq:ck_def}).
\end{proof}

\mypara{64-Bit Separation Above Pettie--Wang}
Non-vacuity in Theorem~\ref{thm:lb} requires $a \ge \ln\widetilde{K}$ (since $\mathcal{J}(a) \le e^a$ and $\mathcal{R}(a) \le 2e^a$ by Lemma~\ref{lem:finite_u}, $a{=}0$ constrains only $\widetilde{I} \le 1$, where $B \ge 2.382\widetilde{I}$ holds vacuously for $B \ge 3$; full width $d = \ln(2^{64}/4) \approx \Nm{d64.val}$ yields width factors \textbf{\Nm{d64.mrg.arb.3}} [Fisher, $\tau{=}1.20$; \Nm{d64.mrg.elem.3} elem.] and \textbf{\Nm{d64.rate.arb.3}} [rate, $\tau{=}0.85$]). On the config-matched windows $[\lambda_{\min}, 2^{62}]$ of Table~\ref{tab:practical_asymp} ($d \in [30.28, 33.22]$, $\lambda_{\min} > 17{,}192 \gg \widetilde{I}$, strictly non-vacuous), 360-bit Arb certifies \textbf{\Nm{mw.lo.3}--\Nm{mw.hi.3}} ($+\Nm{mw.pct.lo}\%$--$+\Nm{mw.pct.hi}\%$ above \Nm{c.pw21.5}) using outward-rounded $c_{\dot{I}}(1.20) \ge$ \Nm{d64.cI.arb.5} and $c_{\dot{R}}(0.85) \ge$ \Nm{d64.cR.arb.5} (\cite[Thm.~6]{pettie2021fish} gives $1{-}\max\{8d^{-1/4}, 5e^{-d/2}\} < 0$ for $d{<}4096$; \eqref{eq:finite_d_lb} hits \Nm{d1e6.mrg.elem.5} at $d{=}10^6$). Steps~1--4 use only independent Poisson cells ($\sum p_i{=}1$ enters only Lemma~\ref{lem:finite_u}(2) and state-change rate).

\subsection{Linearizable Homomorphic Overflow Collapse Cannot Bypass \texorpdfstring{\Nm{c.merge.6}}{2.459261}}
A natural objection asks why a variable-length entropy sketch cannot achieve \Nm{c.pw21.5} inside a fixed $B$-bit buffer simply by collapsing overflow states into the absorbing top state $\top$, preserving strict union homomorphism. Because Theorem~\ref{thm:lb} depends only on $|\mathcal{S}| \le 2^B$ and linearizability of the resulting closure $\Phi_{\mathcal{U}}$ on the original generator support $\mathcal{C}$ (leaving non-linearizable collapses and general join-semilattices to \S\ref{sec:beyond_lin}), any linearizable upward-closed $\top$-collapse on $\mathcal{C}$ is caught by the same bound; by contrast, collapsing arbitrary states whose encoded byte length exceeds $B$ generally creates cyclic cross-cell implications (non-linearizable, falling under \S\ref{sec:beyond_lin}).

\begin{corollary}[Homomorphic $\top$-Collapse Bound]
\label{cor:top_collapse}
Under the hypotheses of Theorem~\ref{thm:lb}, let $\mathcal{S}_0 \subseteq \{0,1\}^N$ be a closure system and let $\mathcal{U} \subseteq \mathcal{S}_0$ be any upward-closed overflow set collapsed to $\top$, forming $\mathcal{S} = (\mathcal{S}_0 \setminus \mathcal{U}) \cup \{\top\}$ with closure $\Phi_{\mathcal{U}}$ and $|\mathcal{S}| \le 2^B$. If $\mathcal{S}$ is linearizable on $\mathcal{C}$ and satisfies $\inf_{x \in [a,b]} \mathcal{J}_{\Phi_{\mathcal{U}}}(x) > 0$, then Theorem~\ref{thm:lb} applies verbatim to $\Phi_{\mathcal{U}}$, forcing for every $\tau > 0$
\[
  \frac{B}{\inf_{x \in [a,b]} \mathcal{J}_{\Phi_{\mathcal{U}}}(x)} \;\ge\; \frac{d}{d+\tau}\,c_{\dot{I}}(\tau),
\]
and at $\tau = \tau(d)$ ($d \ge 3$) giving $\ge \Nm{c.merge.6} - O((\ln\ln d)/d)$.
\end{corollary}

\noindent\textit{Mechanism.} The obstruction is \emph{coordinatewise entropy dominance} inside~\eqref{eq:gibbs_chain}: collapsing the upward cone $\{y \supseteq A\}$ of an $m$-cell subset $A$ on $2m$ equal cells to $\top$ stays acyclic and attains $\log_2|\mathcal{S}|/H_{P_\lambda}(Y) \to 1$ at hit rate $0.5$, yet monotonicity still forces $\Pr_{Q_\lambda}(\phi_k = 0) \ge h_{\pi(k)}(x)$, swapping $\dot{H}$ for $\widetilde{H}$ ($+\Nm{pct.above.pw21.1}\%$).

\begin{remark}[Scope Boundaries]
\label{rem:scope}
First, collapsing rare states into a non-upward-closed \texttt{FAIL} symbol generically breaks union homomorphism: on 2-cell witness $(p_0,p_1)$ with $\Pr[\texttt{FAIL}]{=}8{\times}10^{-13}$, identifying $x_1{=}00$ and $x_2{=}01$ into one \texttt{FAIL} class prevents any commutative, associative, idempotent merge from separating $x_1\vee 10{=}10$ from $x_2\vee 10{=}11$ (0/16 homomorphic binary extensions; absorbing $\top$ falls instead under Cor.~\ref{cor:top_collapse} and \S\ref{sec:beyond_lin}). Second, Theorem~\ref{thm:lb}(b) strictly bounds the worst-window ratio $B/\!\inf_{x\in[a,b]}\mathcal{R}(x)$ (where $\mathcal{R}(x) = 2\lambda\mathbb{E}[p_{\mathrm{chg}}(Y)]$ and $\frac{1}{2}\int_x^{x+\ln 2}\mathcal{R}(u)\,du$ gives expected writes per doubling) and does \emph{not} imply a fixed-space Martingale variance lower bound: in any finite closure with $|\mathcal{S}| \le 2^B$, $\Pr_{P_\lambda}(Y=\top) > 0$ makes $\mathbb{E}[1/\allowbreak p_{\mathrm{chg}}(Y)] = \infty$, while Jensen's inequality ($\mathbb{E}[1/\allowbreak p_{\mathrm{chg}}(Y)] \ge 1/\allowbreak \mathbb{E}[p_{\mathrm{chg}}(Y)]$) makes $1/\allowbreak \mathbb{E}[p_{\mathrm{chg}}(Y)]$ only a lower-bound proxy for $\mathbb{E}[1/\allowbreak p_{\mathrm{chg}}(Y)]$ that also omits accumulator bits.
\end{remark}

\subsection{Beyond Linearizability: General Semilattices}
\label{sec:beyond_lin}
Extending Theorem~\ref{thm:lb} to arbitrary \emph{non-linearizable} join-semilattices (including non-linearizable $\top$-collapses, e.g., collapsing all ${\ge}2$-hit subsets of a 3-cell bitmap to $\top$, which creates the cycle $\{a,b\}\to c,\; \{a,c\}\to b$) remains open, the fixed-capacity analog of the first open problem in~\cite{pettie2021fish}. For small fixed $B$, Fisher data-processing yields an unconditional integral bound:

\begin{proposition}[General Finite Join-Semilattice Bound]
\label{prop:gen_semi}
For any non-constant finite join-semilattice sketch $(\mathcal{S}, \vee)$ (linearizable or not) with $|\mathcal{S}| \le 2^B$ and $\inf_{x \in [a,a+D]} \mathcal{J}(x) > 0$ in the single-cell Poissonized model over window $[a, a+D]$ ($D > 0$),
\[
  \frac{B}{\inf_{x \in [a,a+D]} \mathcal{J}(x)} \;\ge\; \frac{B \cdot D}{(2^B - 1)\, I_0}.
\]
\end{proposition}

\begin{proof}
Merging all cells mapped to each $s \in \mathcal{S}\setminus\{\bot\}$ into weights $w_s \ge 0$ yields at most $2^B - 1$ non-bottom generators. By Fisher data-processing, $\mathcal{J}_Y(x) \le \sum_{s \neq \bot,\, w_s > 0} \dot{I}(w_s e^x)$, and integrating over $[a,a+D]$ gives $D \inf_{[a,a+D]} \mathcal{J}_Y(x) \le (2^B - 1)I_0$.
\end{proof}

To probe small-state behavior empirically, we audited all \Nm{lat.total} unlabeled finite lattices with $2 \le |\mathcal{S}| \le 8$ under join-irreducible support $J(\mathcal{S})$ (\Nm{lat.lin} linearizable vs.\ \Nm{lat.nonlin} non-linearizable via Proposition~\ref{prop:dag}). Heuristic \texttt{scipy} weight searches over $D \in \{1,2,4,8,16,64\ln 2\}$ found \emph{0 of the \Nm{lat.nonlin} non-linearizable lattices beating the best linearizable lattice at the same lattice size and window width}; at $B{=}3$ ($|\mathcal{S}|{=}8$), 360-bit Arb encloses the realized objective of each search-selected candidate at $D \in \{2,4,8,64\ln 2\}$ within $+\Nm{lat.id157.pen}\%$--$+\Nm{lat.id212.pen}\%$ separation (Appendix~D). While heuristic over weights and restricted to $|\mathcal{S}| \le 8$, this search finds no small-state counterexample to linearizable optimality.

%% file: sections/04_cappcsa.tex
\section{Pillar II: Matching Construction \texorpdfstring{\cappcsa{}}{CapPCSA}, Fast Codecs, and Sliding Window}
\label{sec:cappcsa}

\cappcsa{} turns Theorem~\ref{thm:lb}'s variational optimum into a practical deterministic fixed-wire sketch. Conceptually, it behaves like a PCSA bitmap whose floor $L$ advances monotonically whenever any upper layer exceeds a pre-computed Hamming-weight cap $c_m$ or the sparse far tail exceeds $C_{\mathrm{tail}}$ hits. Below $L$, all bits are implicitly $1$; just above $L$, $s$ dense layers sit uncompressed; higher layers stay provably sparse, enabling both an exact combinadic wire encoding (matching \Nm{c.merge.6} asymptotically) and a zero-allocation popcount-orbit rANS codec (\textsc{CapPCSA-Fast}) that runs in microseconds while preserving $\delta{=}0$ and certified sub-$3.00$ inverse-Fisher UBs.

\subsection{Monotone Capped PCSA Closure and Exact Bound}
Fix rational base $q > 1$, column count $g \ge 1$ (Rust supports $K{\le}255$, $n_{\mathrm{rans}}{\le}32$, power-of-two $g{\in}[64, 32{,}768]$, routing 128-bit hashes via integer limits $\lfloor 2^{128} q^{-j}\rfloor$, dropping dummy mass $\lfloor 2^{128} q^{-K}\rfloor / 2^{128}$), level count $K = \lceil \log_q U \rceil$, cutoff $T > \ln 2$, dense-layer count $s = \lfloor \log_q(T/\ln 2) \rfloor + 1 \ge 1$, near-tail cutoff $m_0 \in (s, K]$, non-increasing caps $c_s \ge \dots \ge c_{m_0-1} \ge 0$, and far-tail cap $C_{\mathrm{tail}} \ge 0$. Cells $(r, \ell) \in \{0,\dots,g{-}1\}\times\{1,\dots,K\}$ carry geometric area $p_\ell = \frac{1}{g}(1 - 1/q)q^{-(\ell-1)}$ (residual mass $q^{-K}$ on dummy cell $c_\infty$). Write $N_\ell(X) := \sum_{r=0}^{g-1} X_{r,\ell}$.

\mypara{Higher-Level Cap Closure $\Phi(X)$}
Level $\ell \in \{1,\dots,K\}$ is \emph{triggered} if either
\begin{align}
\label{eq:trigger}
  &\exists\, m \in \{s,\dots,m_0{-}1\}\text{ with }\ell{+}m \le K,\; N_{\ell+m}(X) > c_m, \nonumber \\
  &\quad\text{or}\quad \sum_{j=\ell+m_0}^{K} N_j(X) > C_{\mathrm{tail}}.
\end{align}
Let $L(X) \in \{0,\dots,K\}$ be $0$ if no level triggers, and otherwise the largest triggered level $\ell^*$. Set $(\Phi(X))_{r,\ell} = 1$ for $\ell \le L(X)$ and $(\Phi(X))_{r,\ell} = X_{r,\ell}$ for $\ell > L(X)$. Because $s \ge 1$, triggering level $\ell$ inspects strictly higher levels $j \ge \ell + s \ge \ell + 1$ via monotone counts, ensuring an acyclic implication graph and exact union homomorphism (unlike same-scale $s{=}0$ saturation, which creates intra-level implication cycles).

\begin{figure*}[t]
  \centering
  \includegraphics[width=0.72\textwidth]{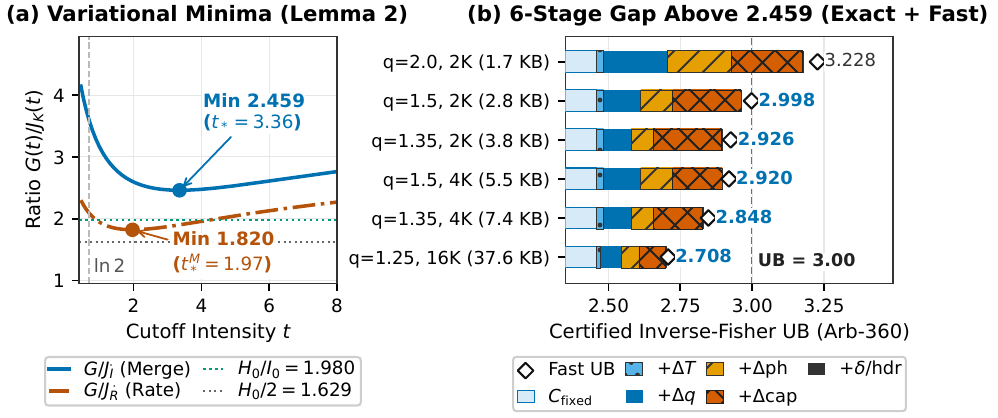}
  \Description{Two-panel plot showing the variational ratio curves attaining unique minima at 2.459 and 1.820, and a horizontal stacked bar chart decomposing the 6-stage gap from 2.459 to certified 64-bit CapPCSA inverse-Fisher upper bounds.}
  \caption{\textbf{(a)} Variational ratios $G(t)/J_K(t)$ attain unique minima \Nm{c.merge.6} ($t_* = \Nm{t.merge.4}$) and \Nm{c.rate.6} ($t_*^{\mathrm{M}} = \Nm{t.rate.4}$) above $\ln 2$ (dotted: entropy floors~\cite{pettie2021fish,pettie2021nonmergeable}). \textbf{(b)} Six-stage gap decomposition to 64-bit \cappcsa{} inverse-Fisher UBs (3-decimal labels round Table~\ref{tab:practical_asymp} upward); five configs certify UB $<3.00$ for \textsc{Exact} and \textsc{Fast}.}
  \label{fig:theory_gap}
\end{figure*}

\begin{theorem}[Tightness, Exact Codec, and Finite-Risk MLE]
\label{thm:cappcsa}
For every valid tuple $(q, g, K, s, m_0, \{c_m\}, C_{\mathrm{tail}})$ with $s \ge 1$:
\begin{enumerate}[leftmargin=*, topsep=1.5pt, itemsep=1.5pt]
  \item \textbf{Exact linearizable closure:} $D(\Phi, \mathcal{C})$ is acyclic in descending level order $\ell = K,\dots,1$, so $\Phi(\Phi(X) \cup \Phi(W)) = \Phi(X \cup W)$ for all $X, W \in \{0,1\}^{g\times K}$ with zero overflow failure ($\delta = 0$).
  \item \textbf{Injective fixed-byte mixed-radix capacity:} Balanced product-tree mixed-radix combinadic serialization (\textsc{CapPCSA-Exact}) encodes every closed state $Y = \Phi(X)$ injectively into a fixed buffer of $B_{\mathrm{bytes}} = \lceil B_{\mathrm{exact}}/8 \rceil$ bytes, where
  \begin{equation}
  \label{eq:bexact}
    B_{\mathrm{exact}} = g s + \left\lceil \log_2\!\left[(K{+}1)\!\left(\prod_{m=s}^{m_0-1}\sum_{u=0}^{c_m}\binom{g}{u}\right)\!\left(\sum_{u=0}^{C_{\mathrm{tail}}}\binom{g(K{-}m_0)}{u}\right)\right]\right\rceil.
  \end{equation}
  \item \textbf{Cylinder-set maximum-likelihood estimation (MLE):} Every closed state $Y$ with floor $L = L(Y)$ has preimage $\{Z : Z_{\cdot,j} = Y_{\cdot,j}\;\forall j > L\}$, giving
  {\sloppy
  $\ell(\lambda) = \sum_{j=L+1}^{K}[N_j\ln(1{-}e^{-p_j\lambda}) - (g{-}N_j)p_j\lambda]$
  with strictly decreasing score ($\sum_{j>L} N_j > 0$)
  $\ell'(\lambda) = \sum_{j=L+1}^{K} \frac{N_j p_j}{e^{p_j\lambda}-1} - \lambda_{\mathrm{rem}}$,
  $\lambda_{\mathrm{rem}} := \sum_{j=L+1}^{K}(g{-}N_j)p_j$.
  Since $s \ge 1$, level $K$ never triggers ($L(Y) \le K{-}s < K$) and $L(Y) > 0 \Rightarrow \sum_{j>L} N_j > 0$. The $[0,U]$-constrained MLE returns $0$ if $L{=}0$ and $\sum_{j>0} N_j{=}0$, $U$ if $\sum_{j>L} N_j{=}g(K{-}L)$ or $\ell'(U) \ge 0$, and else solves $\ell'(\lambda){=}0$ on $(0,U)$ via bracketed Newton--bisection in $\ln\lambda$ ($10^{-13}$ tolerance, summing $\lambda_{\mathrm{rem}}$ directly), guaranteeing finite MSE for all $0 \le n \le U$.\par}
  \item \textbf{Certified continuous-phase information lower bound:} Set $p_k := 1 - e^{-T q^{-k}}$, $c_{\mathrm{end}} := c_{m_0-1}$, and $\mu := g T q^{-m_0}/(1-1/q)$. Every level of intensity $t \le T$ stays unforced with probability $\ge 1 - \delta_{\mathrm{cap}}$, where $\delta_{\mathrm{cap}} \in [0, 1]$ and
  \begin{align}
  \label{eq:deltacap}
    \delta_{\mathrm{cap}} &\le \sum_{k=s}^{m_0-1}\Pr[\mathrm{Bin}(g, p_k) > c_k]
    + \sum_{k=m_0}^{K-1}\Pr[\mathrm{Bin}(g, p_k) > c_{\mathrm{end}}] \nonumber \\
    &\quad + \Pr[\mathrm{Poi}(\mu) > C_{\mathrm{tail}}].
  \end{align}
  For every integer $J_{\max} \ge 0$, every $\lambda \in [\lambda_{\min}, U/4]$ with $\lambda_{\min} := T g / (1 - 1/q)$, and kernel $\kappa \in \{\dot{I}, \dot{R}\}$ ($c_\kappa = 1$ for $\dot{I}$, $2$ for $\dot{R}$, with $\mathcal{K}(\lambda) := \sum_i \kappa(p_i\lambda)h_i(\ln\lambda)$),
  \begin{equation}
  \label{eq:cont_phase_lb}
    \frac{\mathcal{K}(\lambda)}{g} \;\ge\; (1 - \delta_{\mathrm{cap}})\!\left(\inf_{r \in [0,1]} \sum_{j=0}^{J_{\max}} \kappa(T q^{-j-r}) - c_\kappa \frac{(U/4) q^{-K}}{g}\right).
  \end{equation}
  \item \textbf{Asymptotic tightness:} Along $q_j = 1 + 1/j$, $g_j = 2^{\lceil 6\log_2 j\rceil}$, $U_j = \lceil e^{j^2}\rceil$, $K_j = \lceil \log_{q_j} U_j \rceil$, $h_j = \ln q_j$, with $T = t_*$ (resp.\ $t_*^{\mathrm{M}}$), $m_0 = \min\{m > s : T q_j^{-m}/(1-1/q_j) \le h_j\}$, $c_m = \lceil g_j(p_m + h_j^2)\rceil$, and $C_{\mathrm{tail}} = \lceil 2 g_j h_j \rceil$,
  \[
    \frac{B_{\mathrm{exact}}}{\inf_{\lambda \in [\lambda_{\min}, U_j/4]} \mathcal{K}(\lambda)} \;\le\; \frac{G(T)}{J_K(T)} + o(1) \qquad (j \to \infty),
  \]
  matching both $C_{\mathrm{fixed}}^{\mathrm{merge}} = \Nm{c.merge.6}$ and $C_{\mathrm{fixed}}^{\mathrm{rate}} = \Nm{c.rate.6}$.
\end{enumerate}
\end{theorem}

\begin{proof}
(1) Filling levels $\le L(X)$ leaves all counts $N_j$ ($j > L(X)$) and $L(X)$ unchanged, giving idempotence; every minimal implication in~\eqref{eq:trigger} points from $j \ge \ell + s \ge \ell + 1$ to $\ell$, so $D(\Phi, \mathcal{C})$ is acyclic (Proposition~\ref{prop:dag}).
(2) Because level $L{+}1$ did not trigger, dense layers $L{+}1\dots L{+}s$ take $g s$ raw bits, near-tail layers have weights $\le c_m$, and the pooled far tail has weight $\le C_{\mathrm{tail}}$. Enumerative combinadic ranking~\cite{cover1973enumerative} plus balanced product-tree mixed-radix packing yields an injection into $[0, 2^{B_{\mathrm{exact}}})$.
(3) Forcing $\ell$ depends only on levels $> \ell$, so $\Phi(Z){=}Y$ iff $Z$ matches $Y$ on levels $> L(Y)$; strict concavity of $\ell(\lambda)$ (whenever $\sum_{j>L} N_j > 0$) and clamping at $\{0,U\}$ bound $|\hat{\lambda}-n|\le U$.
(4) Level $\ell_0$ ($t_0 \le T$) is forced only if some $\ell \ge \ell_0$ triggers~\eqref{eq:trigger}. Any single-level trigger at $j = \ell{+}m$ ($s \le m < m_0$) has offset $k := j{-}\ell_0 \ge s$ and intensity $t_j \le T q^{-k}$; non-increasing caps give $c_m \ge c_k$ for $k < m_0$ (first sum in~\eqref{eq:deltacap}) and $c_m \ge c_{\mathrm{end}}$ for $k \ge m_0$ (second sum). Any far-tail trigger sums hits at $j \ge \ell_0{+}m_0$, dominated by $\mathrm{Poi}(\mu)$ (third term). Union-bounding gives~\eqref{eq:deltacap}; indexing the first level $\le T$ by phase $r \in [0,1]$ and subtracting omitted $\ell > K$ mass gives~\eqref{eq:cont_phase_lb} for all $J_{\max} \ge 0$.
(5) Along $(q_j, g_j, U_j)$ ($h_j = \ln q_j \sim 1/j$, $K_j = \Theta(j^3)$), Hoeffding bounds each binomial tail in~\eqref{eq:deltacap} by $e^{-2 g_j h_j^4} = e^{-\Omega(j^2)}$ and Poisson Chernoff bounds the far tail by $e^{-\Omega(j^5)}$, giving $\delta_{\mathrm{cap}} = e^{-\Omega(j^2)} \to 0$. By $\sum_{u\le c_m}\binom{g_j}{u} \le 2^{g_j h_2(\min(c_m/g_j,1/2))}$, scaling~\eqref{eq:bexact} by $h_j/g_j$ converts dense ($s h_j \approx \ln(T/\ln 2)$) and near-tail layers into the Riemann upper sum $h_j\sum_{m=0}^{m_0-1}\widetilde{H}(T e^{-mh_j}) + O(h_j^2\ln^2(1/h_j)) \le G(T) + o(1)$, while header and far-tail terms vanish as $O(j^{-2}\ln j)$. Meanwhile $v \mapsto K(T e^{-v})$ has finite total variation $V_K < \infty$, so~\eqref{eq:cont_phase_lb} converges to $J_K(T) - O(1/j)$ uniformly in $r \in [0,1]$ (details in App.~B).
\end{proof}

\input{tables/tab_practical_asymp}

\mypara{64-Bit Certificates and Parameter Selection}
Table~\ref{tab:practical_asymp} certifies~\eqref{eq:cont_phase_lb} in 360-bit Arb~\cite{johansson2017arb} over 4,096 phase subintervals ($U{=}2^{64}$, $\lambda \in [\lambda_{\min}, 2^{62}]$; $\{c_m\}$ in Appendix~D): five configs achieve UBs $<3.00$ for \textsc{Exact} ($+\Nm{mw.gap5.ex.lo}\%$--$+\Nm{mw.gap5.ex.hi}\%$ above matched LBs) and \textsc{Fast} ($+\Nm{mw.gap5.fa.lo}\%$--$+\Nm{mw.gap5.fa.hi}\%$; \texttt{q2p0\_g2048} is $3.18$/$3.23$, $+\Nm{mw.gap.q2.ex}\%$/$+\Nm{mw.gap.q2.fa}\%$). Figure~\ref{fig:theory_gap}(b) decomposes the gap above \Nm{c.merge.6}; we set $T{=}1.25\,t_*$ ($s{\in}\{3,5,7\}$ for $q{\in}\{2.0,1.5,1.35\}$; $T_{\mathrm{jump}}{=}5.26$ at $q{=}1.5$) for $g{\le}4{,}096$ and $T{=}1.18\,t_*$ ($s{=}8$) for \texttt{q1p25\_g16384}.

\subsection{Zero-Alloc Microsecond Codecs: \textsc{CapPCSA-Fast} and \textsc{CapPCSA-RawEF}}
\label{sec:fast_codec}
\textsc{CapPCSA-Exact} uses Karatsuba product-tree BigInt encoding and Knuth-D decoding over $B_{\mathrm{sparse}} := B_{\mathrm{exact}} - g s$ bits (\Nm{race.q15.2k.exact.enc.ms}--\Nm{race.q135.2k.exact.dec.ms}\,ms in M5 warm race at $g{=}2{,}048$), whereas OLAP shuffles and rollups need zero-allocation microsecond serialization. Because any injective fixed-byte codec over $Y{=}\Phi(X)$ preserves mergeability and accuracy, we build two zero-BigInt, zero-alloc ($\delta{=}0$) codecs.

\mypara{Popcount-Orbit Symmetric Nibble-rANS (\textsc{CapPCSA-Fast})}
Standard tabled rANS over $b_{\mathrm{sym}}$-bit symbols with $M$-level frequency quantization requires $O(M)$ decode tables per layer (e.g., 256\,KB across 16 layers at $M{=}2^{14}$), thrashing L1 cache. We exploit column symmetry in every capped near-tail layer $m$ ($N_m \le c_m$): under an independent $\mathrm{Bernoulli}(p_m)$ coding model ($p_m := c_m/g$), all $\binom{4}{v}$ 4-bit nibbles of Hamming weight $v \in \{0,1,2,3,4\}$ share model probability $p_m^v (1{-}p_m)^{4-v}$. Grouping the 16 nibbles into \emph{5 popcount orbits} $v \in \{0..4\}$ assigns one integer frequency $f_v \ge 1$ per orbit ($\sum_{v=0}^4 \binom{4}{v} f_v = M := 2^{14}$), shrinking hot decode-orbit state to \textbf{\Nm{fast.lut.orbit.layer}\,B/layer} (\textbf{\Nm{fast.lut.orbit.min.kb}--\Nm{fast.lut.orbit.max.kb}\,KB} total across rANS layers; \Nm{fast.lut.rans.min.kb}--\Nm{fast.lut.rans.max.kb}\,KB rANS L1 LUTs inside \Nm{fast.static.min.kb}--\Nm{fast.static.max.kb}\,KB shared read-only tables incl.\ 128-bit thresholds and $p_\ell$).

Per near-tail layer, \textsc{CapPCSA-Fast} statically chooses the tightest worst-case bit-cost mode among Raw ($g$ bits, used e.g.\ on the densest near-tail layer of \texttt{q1p25\_g16384}), 4-way interleaved \texttt{u32} nibble-rANS~\cite{duda2013ans} ($L_{\mathrm{ans}} = 2^{24}$, base $b = 2^8$, states $x_k \in [2^{24}, 2^{32})$, 4-byte flush per lane), and Fixed-Slot Elias--Fano (EF)~\cite{elias1974efficient}. Our rANS engine combines: (i)~exact 63-bit Barrett reciprocals ($\lceil 2^{63}/f_v\rceil$) eliminating hardware division in encoding; (ii)~a 4-compare orbit lookup plus 1 \texttt{umulh} nibble decoder avoiding $O(M)$ lookup tables; (iii)~a canonical \texttt{active\_mask} ($\lceil n_{\mathrm{rans}}/8\rceil$ bytes) skipping empty rANS layers in $O(1)$ CPU time while keeping the deterministic fixed slot zero-padded; and (iv)~Fixed-Slot EF (using byte-wise \texttt{tzcnt} high-bit scans) for the far tail and ultra-sparse near-tail layers.

\begin{proposition}[Deterministic $\delta{=}0$ Bound and Asymptotic Optimality of \textsc{CapPCSA-Fast}]
\label{prop:fast_codec}
For every \cappcsa{} tuple with $16 \mid g$ and any rANS parameters $(M, L_{\mathrm{ans}})$ with $M$ a power of two, $M \mid L_{\mathrm{ans}}$, and $L_{\mathrm{ans}} \ge 2^8 M$ (instantiated at $M{=}2^{14}, L_{\mathrm{ans}}{=}2^{24}$ in Rust), let each rANS near-tail layer $m$ use popcount-orbit frequencies $f_{m,0},\dots,f_{m,4} \ge 1$ satisfying $\sum_{v=0}^4 \binom{4}{v} f_{m,v} = M$, cumulative orbit boundaries $c_{\mathrm{off}}[m,v]$, and orbit-aware effective bit costs $a_{m,v}^{\mathrm{eff}} := \log_2(M/f_{m,v}) + \log_2\!\bigl(1 + (c_{\mathrm{off}}[m,v{+}1]{-}f_{m,v})/L_{\mathrm{ans}}\bigr)$ (zero renorm overhead at $v{=}0$). Define the 5-point concave-hull bound
\begin{equation}
\label{eq:rans_bound}
  B^{\mathrm{rANS}}_m := \frac{g}{4}\!\max_{\substack{\sum_v \alpha_v = 1,\; \alpha_v \ge 0 \\ \sum_v v\,\alpha_v \le 4 c_m / g}} \sum_{v=0}^4 \alpha_v a_{m,v}^{\mathrm{eff}}.
\end{equation}
Then: (1)~every closed state $Y = \Phi(X)$ emits $\le \lfloor \sum_m B^{\mathrm{rANS}}_m / 8 \rfloor$ rANS renormalization bytes deterministically ($\delta{=}0$), fitting in $B_{\mathrm{fast}}$ fixed bytes (floor $\lceil\log_{256}(K{+}1)\rceil$\,B, raw bitplanes, Fixed-Slot EF bytes, and when $n_{\mathrm{rans}}{>}0$, active mask $\lceil n_{\mathrm{rans}}/8\rceil$\,B, rANS payload, and four flushed states $4\lceil\log_{256}(2^8 L_{\mathrm{ans}})\rceil$\,B, all-zero if \texttt{active\_mask}{=}0 and canonical terminal $x_k{=}L_{\mathrm{ans}}$ plus zero padding otherwise) with reachability ($L \le K{-}s$, level $L$ triggers if $L{>}0$, and no level $> L$ triggers) verification; (2)~at fixed $(M, L_{\mathrm{ans}})$, encode/decode run in $O(g m_0 + C_{\mathrm{tail}} + K(m_0{-}s) + B_{\mathrm{fast}})$ time on byte streams with \textbf{0 BigInt ops} and, within Rust stack bounds $K{\le}255, n_{\mathrm{rans}}{\le}32, C_{\mathrm{tail}}{\le}512$ (\S4.1), uses fixed stack workspaces (\textbf{0 heap allocations} observed in \texttt{GlobalAlloc} roundtrip and malformed-frame tests of caller-buffer \texttt{*\_into} routines); and (3)~along Theorem~\ref{thm:cappcsa}(5)'s sequence, using the Bernoulli orbit-frequency construction in the proof with scaled resolution $M_j = 2^{\lceil 10\log_2 j\rceil}$ and $L_{\mathrm{ans},j} = 2^8 M_j j^2$, $8 B_{\mathrm{fast}}/B_{\mathrm{exact}} \to 1$, matching $C_{\mathrm{fixed}}^{\mathrm{merge}}$ and $C_{\mathrm{fixed}}^{\mathrm{rate}}$.
\end{proposition}

\begin{proof}
(1)~Set virtual state $V := \prod_{k=0}^3 (x_k / L_{\mathrm{ans}}) \cdot 2^{8 E}$ ($E$ emitted bytes; initially $V{=}1$). Encoding a nibble in orbit $v$ with frequency $f_v$ and starting offset $c \le c_{\mathrm{off}}[v{+}1]{-}f_v$ emits $e \ge 0$ bytes so $x_k' = \lfloor x_k / 2^{8e} \rfloor$ satisfies $x_k' 2^{8e} \le x_k$ and $(L_{\mathrm{ans}}/M)f_v \le x_k' < 2^8(L_{\mathrm{ans}}/M)f_v$ (valid since $M \mid L_{\mathrm{ans}}$; hence $\frac{M}{f_v} x_k' \ge L_{\mathrm{ans}}$ and $L_{\mathrm{ans}}/M \le q = \lfloor x_k'/f_v \rfloor < 2^8 L_{\mathrm{ans}}/M$). Writing $x_k' = q f_v + r$ maps $x_k' \mapsto q M + c + r = \frac{M}{f_v} x_k' + c - r(\frac{M}{f_v}{-}1) \le \frac{M}{f_v} x_k'(1 + (c_{\mathrm{off}}[v{+}1]{-}f_v)/L_{\mathrm{ans}})$, maintaining $L_{\mathrm{ans}} \le q M \le x_k^{\mathrm{new}} < 2^8 L_{\mathrm{ans}}$ and increasing $\log_2 V$ by $\le a_{m,v}^{\mathrm{eff}}$ (since $x_k' 2^{8e} \le x_k$). Summing over any valid layer ($\sum_j v_j \le c_m$) is bounded by~\eqref{eq:rans_bound}; skipped inactive layers add $0$. At termination $x_k \ge L_{\mathrm{ans}}$ gives $8 E \le \log_2 V \le \sum_m B^{\mathrm{rANS}}_m$, proving $\delta{=}0$.
(2)~Active words take $O(g m_0)$ time, closure and reachability take $O(C_{\mathrm{tail}} + K(m_0{-}s))$ time, and reading/writing the wire frame (Fixed-Slot EF and canonical padding) takes $O(B_{\mathrm{fast}})$ byte operations.
(3)~Along Theorem~\ref{thm:cappcsa}(5)'s sequence, every near-tail layer with $p_m' := c_m/g_j \le 1/2$ has $p_m' \ge h_j^2 = \Omega(j^{-2})$, so $M_j(p_m')^4 = \Omega(j^2)$; flooring $\bar{f}_{m,v} = \lfloor M_j(p_m')^v(1{-}p_m')^{4-v}\rfloor$ and assigning the remaining mass $M_j - \sum_{v=0}^4 \binom{4}{v}\bar{f}_{m,v} \in [0, 16)$ to orbit $0$ ($f_{m,v}{=}\bar{f}_{m,v}$ for $v{>}0$, $f_{m,0}{=}\bar{f}_{m,0}{+}\text{rem}$) preserves $\sum_v \binom{4}{v} f_{m,v} = M_j$ and gives $a_{m,v}^{\mathrm{eff}} \le v\log_2(1/p_m') + (4{-}v)\log_2(1/(1{-}p_m')) + O(j^{-2})$, yielding $B_m^{\mathrm{rANS}} \le g_j h_2(p_m') + O(g_j j^{-2})$ (layers $p_m' > 1/2$ take $g_j$ raw bits). Across $m_0 = O(j\ln j)$ layers, normalized ($h_j/g_j$) excess rANS bits add $O(j^{-2}\ln j)$, far-tail EF adds $O(h_j^2\log(K_j/h_j))$, and headers/flushes add $O(j\ln j)/(g_j/h_j)$, all $o(1)$. With App.~B ($h_j{:=}\ln q_j$), $h_j B / g_j \le G(T) + o(1)$ for $B \in \{8 B_{\mathrm{fast}}, B_{\mathrm{exact}}\}$, while Theorems~\ref{thm:lb} and~\ref{thm:cappcsa}(4) give $h_j B / g_j \ge G(T) - o(1)$, yielding $8 B_{\mathrm{fast}}/B_{\mathrm{exact}} \to 1$.
\end{proof}

In Table~\ref{tab:practical_asymp} (Rust/Arb byte-exact; the five inverse-Fisher UB $<3.00$ configs sit $+\Nm{mw.gap5.ex.lo}\%$--$+\Nm{mw.gap5.ex.hi}\%$ \textsc{Exact} / $+\Nm{mw.gap5.fa.lo}\%$--$+\Nm{mw.gap5.fa.hi}\%$ \textsc{Fast} above LBs), \textsc{Fast} adds $+\Nm{fast.ovh.min}\%$--$+\Nm{fast.ovh.max}\%$ wire bytes at \Nm{fast.spd.min}$\times$--\Nm{fast.spd.max}$\times$ warm M5 speedup (\Nm{race.q15.2k.fast.enc}--\Nm{race.q135.2k.fast.dec}\,$\mu$s at $g{=}2{,}048$ vs.\ \Nm{ops.cap15.2k.fast.enc}--\Nm{ops.cap135.2k.fast.dec}\,$\mu$s unified; Xeon \Nm{xeon.spd.min}$\times$--\Nm{xeon.spd.max}$\times$). Table-free \textsc{RawEF} preserves $\delta{=}0$ in \Nm{race.q15.2k.ef.enc}--\Nm{race.q135.2k.ef.dec}\,$\mu$s warm (Arb-360 UB \Nm{rawef.q15.2k.ub3} at $q1.5/2$K, \Nm{rawef.q135.4k.ub3}/\Nm{rawef.q125.16k.ub3} at $q1.35/4$K, $q1.25/16$K).

\subsection{Compact Sliding Window and Lazy 128-Bit Hashing}
\label{sec:sliding}
A naive dense $g \times K$ active bitmatrix (incl.\ 8\,B/level counters) consumes \Nm{cap.q135.2k.dense}\,B for a \Nm{cap.q135.2k.wire}\,B \textsc{Exact} wire sketch (\Nm{cap.q125.16k.dense}\,B at $g{=}16{,}384$). We cut this working-RAM overhead \Nm{cap.q135.2k.ramcut}$\times$ (to \Nm{cap.q135.2k.ram}\,B) without altering a single bit of $\Phi(X)$: levels $\le L$ are implicitly $1$, levels $L{+}1\dots L{+}m_0$ live in an $m_0$-layer circular bitplane ring ($m_0 g / 8 + 2 m_0$\,B), and hits $> L{+}m_0$ live in a sorted deduplicated vector \texttt{far\_tail} pre-allocated to capacity $\ge C_{\mathrm{tail}} + 1$ ($4(C_{\mathrm{tail}}{+}1)$\,B). To accelerate streaming inserts, \emph{lazy two-stage hashing} checks the upper 64-bit hash prefix against $\theta_{\mathrm{hi}} = \min(2^{64}{-}1, \lfloor 2^{64} q^{-L}\rfloor)$ first and skips the remaining two 64-bit hash mixes whenever $h_{\mathrm{hi}} > \theta_{\mathrm{hi}}$ (fast-rejecting $\approx 1 - q^{-L}$ of keys once $L>0$; prefix ties fall through to the full 128-bit comparison), preserving bit-exact equivalence with 192-bit hashing. Algorithm~\ref{alg:cappcsa} gives the complete self-contained update, closure, and zero-alloc merge procedures.

\begin{algorithm}[t]
\caption{Compact Sliding-Window \cappcsa{} Update, Closure \& Merge}
\label{alg:cappcsa}
\footnotesize
\begin{algorithmic}[1]
\State \textbf{State:} floor $L$, ring head \texttt{head} ($0$ init), \texttt{win[$m_0$][$g/64$]}, \texttt{cnt[$m_0$]}, sorted \texttt{far\_tail} ($\mathrm{cap}\ge C_{\mathrm{tail}}{+}1$), $\theta_{\mathrm{hi}} = \min(2^{64}{-}1, \lfloor 2^{64} q^{-L}\rfloor)$ (shared hash seed across merged shards)
\Procedure{Close}{$L_{\mathrm{base}}, \texttt{tail\_slice}$}
  \State tally counts $N_j$ for $j \in (L_{\mathrm{base}}, K]$ from \texttt{win} and \texttt{tail\_slice} on stack ($K{\le}255$)
  \State scan $\ell = K{-}1$ down to $L_{\mathrm{base}}{+}1$ tracking suffix sum $S = \sum_{j>\ell+m_0-1} N_j$ to find largest triggered $L^*$ via~\eqref{eq:trigger} (else $L^* \gets L_{\mathrm{base}}$)
  \State advance \texttt{head} by $L^*{-}L_{\mathrm{base}}$ (zero retired planes/counts; reset \texttt{head}${\gets}0$ if $L^*{-}L_{\mathrm{base}}{\ge}m_0$), promote \texttt{tail\_slice} in $(L^*, L^*{+}m_0]$ into \texttt{win} (updating \texttt{cnt}), compact $>L^*{+}m_0$ into \texttt{far\_tail}, set $L \gets L^*$ and $\theta_{\mathrm{hi}}$
\EndProcedure
\Procedure{UpdateLazy}{$k$}
  \State $h_{\mathrm{hi}} \gets \mathrm{Hash64}_1(k)$;\; \If{$h_{\mathrm{hi}} > \theta_{\mathrm{hi}}$} \Return \EndIf \Comment{at or below floor $L$ ($j{\le}L$)}
  \State $(w_{\mathrm{col}}, h_{\mathrm{lo}}) \gets \mathrm{Hash128}_{\mathrm{rest}}(k)$;\; $r \gets w_{\mathrm{col}} \bmod g$;\; if $L{>}0$ and $(h_{\mathrm{hi}},h_{\mathrm{lo}}) \ge \lfloor 2^{128} q^{-L}\rfloor$ \Return;\; $j \gets \mathrm{Level128}(h_{\mathrm{hi}}, h_{\mathrm{lo}})$ (return on dummy $j{>}K$)
  \If{$j \le L + m_0$}
    \State $\mathrm{off} \gets j - (L{+}1)$;\; $p \gets (\texttt{head}{+}\mathrm{off}) \bmod m_0$;\; set bit $r$ in \texttt{win[$p$]}
    \If{bit flipped} \texttt{++cnt[$p$]}; \If{$\mathrm{off} \ge s$ \textbf{and} \texttt{cnt[$p$]} $> c_{\mathrm{off}}$} \Call{Close}{$L, \texttt{far\_tail}$} \EndIf \EndIf
  \Else
    \State insert unique $(j, r)$ sorted into \texttt{far\_tail}; \If{inserted \textbf{and} ($|\texttt{far\_tail}| > C_{\mathrm{tail}}$ \textbf{or} $N_j > c_{m_0-1}$)} \Call{Close}{$L, \texttt{far\_tail}$} \EndIf
  \EndIf
\EndProcedure
\Procedure{MergeInPlace}{\texttt{other}}
  \State $L_{\mathrm{base}} \gets \max(L, \texttt{other}.L)$;\; $\Delta \gets L_{\mathrm{base}}{-}L$;\; zero first $\min(\Delta,m_0)$ retired planes/counts from \texttt{head}, rotate \texttt{head} by $\Delta \bmod m_0$ (reset to $0$ if $\Delta{\ge}m_0$);\; $L \gets L_{\mathrm{base}}$;\; drop covered tail hits and promote $(L, L{+}m_0]$ from \texttt{self.far\_tail} updating \texttt{cnt}
  \State bitwise-OR overlapping \texttt{other.win} planes (recounting \texttt{cnt}) and insert \texttt{other.far\_tail} in $(L_{\mathrm{base}}, L_{\mathrm{base}}{+}m_0]$ into \texttt{win} (incrementing \texttt{cnt} on each flip)
  \State merge-deduplicate both \texttt{far\_tail} slices ($j > L_{\mathrm{base}}{+}m_0$) into \texttt{tmp} (stack[512] if total $\le 512$, else heap);\; \Call{Close}{$L_{\mathrm{base}}, \texttt{tmp}$}
\EndProcedure
\end{algorithmic}
\end{algorithm}

\begin{proposition}[Compact Window \& Zero-Alloc Operations]
\label{prop:sliding}
Under Algorithm~\ref{alg:cappcsa}, the circular $m_0$-layer ring and pre-allocated \texttt{far\_tail} ($\mathrm{cap} \ge C_{\mathrm{tail}}{+}1$, preserved across construction, \texttt{clone()}, \texttt{clear()}, and \texttt{try\_decode\_*}) are bit-identical to lazily closing the dense $g \times K$ bitmatrix. Whenever $K \le 255$, $n_{\mathrm{rans}} \le 32$, and $2C_{\mathrm{tail}} \le 512$ (all six 64-bit configs, $C_{\mathrm{tail}} \le 180$), \textsc{MergeInPlace} uses only fixed stack workspaces (\texttt{tmp[512]}) with 0 extra heap, and \textsc{MergeInPlace} plus \textsc{Fast}/\textsc{RawEF} \texttt{*\_into} routines recorded \textbf{0 heap allocations} in Rust \texttt{GlobalAlloc} tests (merges, roundtrips, malformed frames).
\end{proposition}

\begin{proof}
By Theorem~\ref{thm:cappcsa}(1), eager closure preserves semantics. On a closed state at $L$, hits at $j \le L{+}s$ cannot trigger $\ell > L$; hits in $(L{+}s, L{+}m_0]$ trigger only if $N_j > c_{\mathrm{off}}$; hits at $j > L{+}m_0$ trigger only if $|\texttt{far\_tail}| > C_{\mathrm{tail}}$ or $N_j > c_{m_0-1}$. In \textsc{MergeInPlace}, advancing \texttt{self} to $L_{\mathrm{base}}$ prevents aliasing, using stack \texttt{tmp[512]} with 0 heap allocations whenever $2C_{\mathrm{tail}} \le 512$ (all six 64b configs). Both \texttt{try\_decode\_*\_into} routines validate bounds, reachability, and canonical padding with 0 heap allocs (\texttt{GlobalAlloc}) and 0 drift over 2,000+ fuzz frames; \texttt{try\_merge\_inplace} requires matching $(q,g)$.
\end{proof}

Table~\ref{tab:practical_asymp} confirms \Nm{ramcut.min}$\times$--\Nm{ramcut.max}$\times$ active RAM reduction (\Nm{cap.q135.2k.dense}\,B to \Nm{cap.q135.2k.ram}\,B at $q{=}1.35, 2$K), verified across dense bitmatrices, \textsc{Exact}/\textsc{Fast}/\textsc{RawEF} roundtrip checks, and bit-identical 32-shard trees with $0$ mismatches.

\begin{remark}[Structural Constraints on Anchored-Register Closures]
\label{rem:anchored}
In a fixed-width $k$-th-largest shared-anchor register array (our $k$-anchor staircase closure inspired by Curtain~\cite{pettie2021nonmergeable}), $g$ buckets share anchor level $A$, retain an $R$-level window above $L = \max(1, A{-}R{+}1)$, and store $d(\Delta) \in \{0,\dots,R{-}1{-}\Delta\}$ history bits at offset $\Delta = A - u_r$ (and $d_{\mathrm{ov}}$ bits above $A$). Appendix~C provides a proof sketch and exhaustive verification ($0$ mismatches across $g{=}3, k{=}2$ for $R{=}3, K{=}5$ [$30$ schedules / \Nm{thm3.trans} transitions], $R{=}2, K{=}3$ [$8$ schedules / $2{,}097{,}152$ pairs]) that exact union homomorphism holds under sufficiency conditions $0 \le d(\Delta{-}1){-}d(\Delta) \le 1$, $d_{\mathrm{ov}} \ge d(0)$, and ($d(0){<}R{-}1 \Rightarrow d_{\mathrm{ov}} \le d(0){+}1$), which are necessary when $2 \le k \le g{-}1$ and $K \ge R{+}1$. Original adjacent-difference Curtain~\cite{pettie2021nonmergeable} is linearizable with fixed 3-bit columns, attaining smoothed Fisher MVP \Nm{curtain.3b.mvp} (family best \Nm{curtain.int.mvp}--\Nm{curtain.3b.mvp}); our $k$-anchor Staircase-6b ($B{=}6$ theory \Nm{st.b6.slot.lo}--\Nm{st.b6.worst.hi}) measures \Nm{ph.b16.curt.mean2} 16-phase (\Nm{s1m.curt4k.mvp2}--\Nm{s1m.curt8k.mvp2} at $n{=}10^6$), whereas pooling tail capacity jointly across columns lets \cappcsa{} break below $3.00$ toward \Nm{c.merge.6}.
\end{remark}

%% file: tables/tab_practical_asymp.tex
\begin{table*}[t]
\centering
\caption{\textbf{Five 64-bit \cappcsa{} configs certify worst-phase inverse-Fisher UBs $<3.00$ (\textsc{Exact} $+\Nm{mw.gap5.ex.lo}\%$--$+\Nm{mw.gap5.ex.hi}\%$, \textsc{Fast} $+\Nm{mw.gap5.fa.lo}\%$--$+\Nm{mw.gap5.fa.hi}\%$ above Arb-360 $[\lambda_{\min}, 2^{62}]$ LBs \Nm{mw.lo.3}--\Nm{mw.hi.3}) and cut RAM \Nm{ramcut.min}$\times$--\Nm{ramcut.max}$\times$.} \textsc{Fast} adds $+\Nm{fast.ovh.min}\%$--$+\Nm{fast.ovh.max}\%$ bytes at \Nm{fast.spd.min}$\times$--\Nm{fast.spd.max}$\times$ M5 (\Nm{xeon.spd.min}$\times$--\Nm{xeon.spd.max}$\times$ Xeon) speedup ($\lceil\lambda_{\min}\rceil$; $\lfloor\mathrm{LB}\rfloor$; \textsc{RawEF}: $\delta{=}0$, UB $<3.00$ for $g{\ge}4$K at $q{\le}1.35$; App.~D).}
\label{tab:practical_asymp}
\footnotesize
\tblclean
\renewcommand{\arraystretch}{1.00}
\setlength{\tabcolsep}{3.0pt}
\begin{tabular}{@{}l r r r r c r r r r@{}}
\toprule
\textbf{Config} ($q, g$) & \textbf{Exact (B)} $\downarrow$ & \textbf{Fast (B)} $\downarrow$ & \textbf{RAM (B)} & \textbf{vs.\ Dense} & $(s, m_0, C_{\mathrm{tail}})$ & $\lambda_{\min}$ & \textbf{Match LB} $\uparrow$ & \textbf{Exact UB} $\downarrow$ & \textbf{Fast UB} $\downarrow$ \\
\midrule
\rowhl \texttt{q2p0\_g2048} ($2.00$, $2{,}048$) & 1{,}699 & 1{,}724 & 2{,}720 & 6.21$\times$ & (3, 10, 34) & 17{,}193 & 2.363 & 3.18062 & 3.22743 \\
\rowhl \texttt{q1p5\_g2048} ($1.50$, $2{,}048$) & 2{,}812 & 2{,}843 & 4{,}788 & 6.07$\times$ & (5, 18, 35) & 25{,}790 & 2.362 & \textbf{2.96512} & \textbf{2.99781} \\
\rowhl \texttt{q1p35\_g2048} ($1.35$, $2{,}048$) & 3{,}757 & 3{,}793 & 6{,}602 & 5.92$\times$ & (7, 25, 37) & 33{,}158 & 2.362 & \textbf{2.89749} & \textbf{2.92525} \\
\rowhl \texttt{q1p5\_g4096} ($1.50$, $4{,}096$) & 5{,}498 & 5{,}537 & 9{,}492 & 6.03$\times$ & (5, 18, 59) & 51{,}579 & 2.361 & \textbf{2.89884} & \textbf{2.91941} \\
\rowhl \texttt{q1p35\_g4096} ($1.35$, $4{,}096$) & 7{,}341 & 7{,}385 & 13{,}102 & 5.87$\times$ & (7, 25, 62) & 66{,}315 & 2.360 & \textbf{2.83090} & \textbf{2.84787} \\
\rowhl \texttt{q1p25\_g16384} ($1.25$, $16{,}384$) & 37{,}480 & 37{,}568 & 72{,}474 & 5.65$\times$ & (8, 35, 180) & 324{,}599 & 2.355 & \textbf{2.70116} & \textbf{2.70750} \\
\midrule
\multicolumn{10}{@{}l}{\rule{0pt}{2.3ex}\textit{Large-$g$ Tightening Milestones (360-bit Arb Eq.~\eqref{eq:bexact}; gap above $C_{\mathrm{fixed}}^{\mathrm{merge}}{=}\Nm{c.merge.5}$ / $C_{\mathrm{fixed}}^{\mathrm{rate}}{=}\Nm{c.rate.5}$):}} \\[1pt]
\midrule
\textbf{Milestone (Thm.~\ref{thm:cappcsa})} & \textbf{Mode} & $q$ & $g$ & $K$ & \multicolumn{2}{r}{\textbf{Wire Capacity (bits)}} & & \textbf{Ratio UB} & \textbf{Gap ($+$)} \\
\midrule
\texttt{Merge\_q1.5\_g1e6} & Merge & $3/2$ & $10^{6}$ & 110 & \multicolumn{2}{r}{10{,}155{,}466} & & 2.739286 & $+0.280025$ \\
\texttt{Merge\_q1.05\_g1e8} & Merge & $21/20$ & $10^{8}$ & 910 & \multicolumn{2}{r}{7{,}650{,}759{,}274} & & 2.493872 & $+0.034611$ \\
\texttt{Merge\_q1.005\_g1e11} & Merge & $201/200$ & $10^{11}$ & 8{,}895 & \multicolumn{2}{r}{73{,}563{,}271{,}439{,}667} & & \textbf{2.462862} & $+0.003601$ \\
\texttt{Rate\_q1.05\_g1e8} & Rate & $21/20$ & $10^{8}$ & 910 & \multicolumn{2}{r}{6{,}558{,}174{,}167} & & 1.850363 & $+0.029935$ \\
\texttt{Rate\_q1.005\_g1e11} & Rate & $201/200$ & $10^{11}$ & 8{,}895 & \multicolumn{2}{r}{62{,}875{,}914{,}196{,}270} & & \textbf{1.823523} & $+0.003095$ \\
\bottomrule
\end{tabular}
\end{table*}

%% file: sections/05_ell_divergence.tex
\section{Pillar III: Divergence of Continuous Smoothed ExaLogLog}
\label{sec:ell}

Independent fixed-width register sketches such as HyperLogLog, UltraLogLog, and ExaLogLog avoid cross-column encoding entirely, operating in place in RAM. To explain why even the optimal continuous smoothed ELL relaxation stays at \Nm{ell.fstar.64.2} on 64-bit universes and cannot approach any universal constant as $U \to \infty$, we establish the exact two-term divergence law for a continuous relaxation of Ertl's smoothed ELL model~\cite[Eq.~(3) \& \S2.4]{ertl2025ell}.

\begin{figure}[t]
  \centering
  \includegraphics[width=0.78\columnwidth]{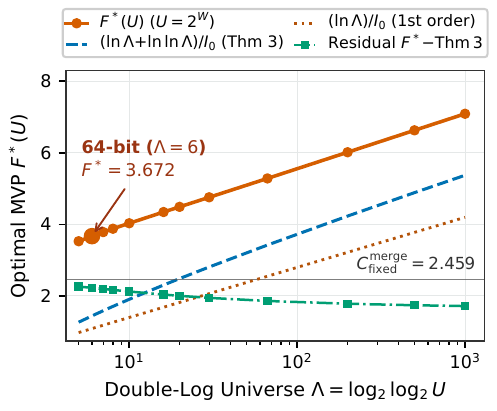}
  \Description{Line plot comparing numerically minimized smoothed ExaLogLog MVP against Theorem 3's two-term formula and first-order logarithmic term across universe sizes from 32 bits to 1e300 bits, showing an O(1) residual.}
  \caption{Numerically minimized smoothed ELL MVP $F^*(U)$ ($U{=}2^W$) vs.\ Theorem~\ref{thm:ell_div}'s $\frac{1}{I_0}(\ln\Lambda + \ln\ln\Lambda)$ ($\Lambda{=}\log_2\log_2 U$) across $W \in [32, 10^{300}]$: residual stabilizes to $O(1)$ (\Nm{ell.res2.32} $\to$ \Nm{ell.res2.1e300}), tracking 64-bit smoothed ELL's \Nm{ell.fstar.64} (two-term \Nm{ell.twoterm.64} + $O(1)$ residual \Nm{ell.res2.64}).}
  \label{fig:ell_div}
\end{figure}

\input{tables/tab_showdown_main}

\begin{theorem}[Two-Term Divergence of Continuous Smoothed ELL]
\label{thm:ell_div}
In the continuous relaxation of Ertl's smoothed $\mathrm{ELL}(b, d_{\mathrm{ell}})$ model covering universe $U = 2^W$ ($W = \log_2 U$, $M := W \ln 2$, $\Lambda := \log_2 W = \log_2\log_2 U$), with $x = \ln b > 0$, $q(x) = \max(1, \log_2(M/x))$, $d_{\mathrm{ell}} \ge 0$, $c = e^{-d_{\mathrm{ell}} x}/(e^x - 1) \in (0, (e^x - 1)^{-1}]$, and Hurwitz zeta function $z(c) = \zeta(2, 1{+}c) = \sum_{k=1}^\infty (k+c)^{-2}$, the optimal smoothed MVP
\begin{equation}
\label{eq:ell_opt}
  F(x, d_{\mathrm{ell}}) := \frac{\bigl(q(x) + d_{\mathrm{ell}}\bigr)\,x}{z(c)}, \qquad F^*(U) := \inf_{x > 0,\; d_{\mathrm{ell}} \ge 0} F(x, d_{\mathrm{ell}})
\end{equation}
satisfies both upper and lower bounds:
\begin{equation}
\label{eq:ell_div}
  F^*(U) \;=\; \frac{1}{I_0}\Bigl[\ln\Lambda + \ln\ln\Lambda\Bigr] + O(1) \qquad (U \to \infty).
\end{equation}
\end{theorem}

\begin{proof}
Assume $W$ is large enough that $M = W\ln 2 \ge 4$ and $\Lambda = \log_2 W \ge 2$.

\emph{Step 1 (Exclusion of $x \ge 1$).} Since $d_{\mathrm{ell}} \ge 0$ and $0 < z(c) < z(0) = I_0$, we have $F(x,d_{\mathrm{ell}}) \ge x q(x)/I_0$. For $1 \le x \le \sqrt{M}$, $q(x) \ge \frac{1}{2}\Lambda + O(1)$, so $F \ge \frac{1}{2I_0}\Lambda - O(1)$; for $x > \sqrt{M}$, $q(x) \ge 1$ gives $F > \sqrt{M}/I_0 = \Omega(2^{\Lambda/2}) \gg \ln\Lambda$.

\emph{Step 2 (Uniform infimum on $0 < x \le 1$).} Here $q(x) = \log_2(M/x)$ is unclipped, and substituting $d_{\mathrm{ell}} x = -\ln c - \ln(e^x - 1)$ rewrites~\eqref{eq:ell_opt} as
\[
  F(x, c) = \frac{A(x) - \ln c}{z(c)}, \qquad A(x) := x \log_2(M/x) - \ln(e^x - 1).
\]
Uniformly on $(0,1]$, $|A(x) - (\Lambda x - \ln x)| < 2$, so minimizing $\Lambda x - \ln x$ at $x = 1/\Lambda$ yields $A_* := \inf_{x \in (0,1]} A(x) = \ln\Lambda + O(1)$.

\emph{Step 3 (Exclusion of $c \ge 1$ on $0 < x \le 1$).} Write $V(c) := A(x) - \ln c > 0$. Integral comparison gives $z(c) \le 1/c$ and $-z'(c) \ge 1/(c+1)^2$. Whenever $V(c) \ge 4$, differentiating yields $\partial_{c'} F \ge (-1/(c')^2 + 4/(c'+1)^2)/z(c')^2 \ge 0$ on $[1,c]$, so $F(x,c) \ge A_*/z(1) = \frac{1}{I_0 - 1}\ln\Lambda - O(1)$, which exceeds~\eqref{eq:ell_div} by a factor $I_0/(I_0-1) \approx 2.55$. If instead $V(c) < 4$, then $d_{\mathrm{ell}} x < 4 \implies cx > e^{-4}/(e-1)$, giving $F \ge c x q(x) = \Omega(\Lambda)$.

\emph{Step 4 (Global lower bound on $0 < x \le 1, 0 < c < 1$).} Since $z''(c) = 6\zeta(4, 1{+}c) > 0$, convexity gives the right-endpoint slope bound $-z'(c) \ge -z'(1) = 2\zeta(3,2)$ on $[0,1]$, hence $z(c) \le I_0 - 2\zeta(3,2)\,c$ and $1/z(c) \ge 1/I_0 + \kappa c$ with $\kappa := 2\zeta(3,2)/I_0^2 > 0$. Therefore,
\[
  F(x, d_{\mathrm{ell}}) \;\ge\; \frac{A_*}{I_0} + \frac{\ln(1/c)}{I_0} + \kappa A_* c \;\ge\; \frac{1}{I_0}\bigl[\ln\Lambda + \ln\ln\Lambda\bigr] - O(1),
\]
where the minimum over $c > 0$ is attained at $c_0 = 1/(I_0 \kappa A_*)$.

\emph{Step 5 (Matching upper bound).} Setting $x = 1/\Lambda \in (0,1]$ and $c = 1/\ln\Lambda \in (0,1)$ (feasible for all $\Lambda > e$) gives $z(c) = I_0 - O(1/\ln\Lambda)$ and $A(1/\Lambda) - \ln c = \ln\Lambda + \ln\ln\Lambda + O(1)$, yielding $F(1/\Lambda, d_{\mathrm{ell}}) = \frac{1}{I_0}[\ln\Lambda + \ln\ln\Lambda] + O(1)$.
\end{proof}

Figure~\ref{fig:ell_div} confirms Theorem~\ref{thm:ell_div} numerically from $W=32$ ($\Lambda=5$) to $W=10^{300}$ ($\Lambda\approx 997$): $F^*(2^W)$ rises from \Nm{ell.fstar.64} at 64-bit to \Nm{ell.fstar.1e300} at $W=10^{300}$, with residual $O(1)$ in [\Nm{ell.res2.1e300.lo}, \Nm{ell.res2.32.hi}] (\Nm{ell.res2.64} at 64-bit). Independent registers must pay $q(x) = \log_2((W\ln 2)/x)$ rank bits on $x \le 1$, whereas \cappcsa{} amortizes one floor $L$ over all $g$ columns to approach \Nm{c.merge.6} along Theorem~\ref{thm:cappcsa}(5).

%% file: tables/tab_showdown_main.tex
\begin{table*}[t]
\centering
\caption{\textbf{\cappcsa{}-Fast hits \Nm{ph.b16.fast135.mean2}--\Nm{ph.b16.fast15.mean2} 16-phase mean MVP (\Nm{s1m.fast15.2k.mvp2}--\Nm{s1m.fast135.2k.mvp2} at $n{=}10^6$), beating ELL-28b/32b by \Nm{ph.cut.fast15.ell28}\%--\Nm{ph.cut.fast135.ell32}\% and Staircase-6b by \Nm{ph.cut.fast15.curt}\%--\Nm{ph.cut.fast135.curt}\%.} Apple M5 (4k paired seeds at $n{=}10^6$, 2k/ph 16-ph; \texttt{zstd}: 400-seed B $\times$ 4k RelMSE; \textsc{Fast} excl.\ \Nm{fast.static.min.kb}--\Nm{fast.static.max.kb}\,KB LUTs; CPC: live/peak). Bold: best fixed-wire.}
\label{tab:showdown_main}
\footnotesize
\tblclean
\renewcommand{\arraystretch}{1.00}
\setlength{\tabcolsep}{3.0pt}
\begin{tabular}{@{}l l r r r r r r r@{}}
\toprule
\textbf{Sketch} & \textbf{Wire} & \textbf{Wire (B)} $\downarrow$ & \textbf{Work RAM (B)} $\downarrow$ & \textbf{16-Ph Mean/Max} $\downarrow$ & \textbf{Wire MVP@$10^6$ ($\pm$SE)} $\downarrow$ & \textbf{RelRMSE} $\downarrow$ & \textbf{P99 Err} $\downarrow$ & \textbf{RAM MVP} $\downarrow$ \\
\midrule
\multicolumn{9}{@{}l}{\textit{Tier 1: Matched Fixed Wire $\le 4{,}096$\,B (plus variable-wire \texttt{libzstd} and Apache CPC \texttt{lgK}{=}12 references)}} \\
\midrule
\rowhl \cappcsa{}-Exact ($q{=}1.5, 2$K) & Fixed & 2{,}812 & 4{,}788 & -- & \textbf{2.6252} $\pm$ 0.0577 & 1.080\% & 2.80\% & 4.470 \\
\rowhl \textbf{\cappcsa{}-Fast ($q{=}1.5, 2$K)} & Fixed & 2{,}843 & 4{,}788 & 2.83 / 2.99 & 2.6542 $\pm$ 0.0583 & 1.080\% & 2.80\% & 4.470 \\
\rowhl \cappcsa{}-RawEF ($q{=}1.5, 2$K) & Fixed & 2{,}989 & 4{,}788 & -- & 2.7905 $\pm$ 0.0613 & 1.080\% & 2.80\% & 4.470 \\
\rowhl \textbf{\cappcsa{}-Fast ($q{=}1.35, 2$K)} & Fixed & 3{,}793 & 6{,}602 & \textbf{2.79} / \textbf{2.91} & 2.7226 $\pm$ 0.0598 & \textbf{0.947\%} & \textbf{2.40\%} & 4.739 \\
\rowhl \cappcsa{}-RawEF ($q{=}1.35, 2$K) & Fixed & 3{,}990 & 6{,}602 & -- & 2.8640 $\pm$ 0.0629 & \textbf{0.947\%} & \textbf{2.40\%} & 4.739 \\
Staircase-6b (ours, $4$K) & Fixed & 3{,}424 & 8{,}192 & 3.28 / 3.42 & 3.1825 $\pm$ 0.0712 & 1.078\% & 2.77\% & 7.614 \\
Orig-ELL-28b ($m{=}1024$) & Fixed & 3{,}584 & 4{,}096 & 3.70 / 3.92 & 3.7247 $\pm$ 0.0838 & 1.140\% & 2.94\% & 4.257 \\
Orig-ELL-32b ($m{=}1024$) & Fixed & 4{,}096 & 4{,}096 & 3.79 / 3.94 & 3.7815 $\pm$ 0.0848 & 1.074\% & 2.80\% & \textbf{3.781} \\
ULL-8b ($m{=}4096$) & Fixed & 4{,}096 & 4{,}096 & 4.57 / 4.79 & 4.5948 $\pm$ 0.1037 & 1.184\% & 3.10\% & 4.595 \\
Dense-HLL-6b ($m{=}4096$) & Fixed & 3{,}072 & 4{,}096 & -- & 6.3940 $\pm$ 0.1392 & 1.613\% & 4.06\% & 8.525 \\
Orig-ELL-32b + \texttt{zstd} & Var & 3{,}272 (max 3{,}319) & 4{,}096 & -- & 3.0212 (var) & 1.074\% & 2.80\% & 3.781 \\
ULL-8b + \texttt{zstd} & Var & 2{,}235 (max 2{,}258) & 4{,}096 & -- & 2.5072 (var) & 1.184\% & 3.10\% & 4.595 \\
Apache CPC (\texttt{lgK}{=}12, ICON) & Var & 2{,}489 (emp 2{,}580) & 5{,}120 / 37{,}888 & 2.33 / 2.45 & 2.3669 $\pm$ 0.0517 & 1.090\% & 2.73\% & 4.87 / 36.0 \\
\midrule
\multicolumn{9}{@{}l}{\textit{Tier 2: Matched Fixed Wire $\le 8{,}192$\,B (plus variable-wire \texttt{libzstd} and Apache CPC \texttt{lgK}{=}13 references)}} \\
\midrule
\rowhl \cappcsa{}-Exact ($q{=}1.35, 4$K) & Fixed & 7{,}341 & 13{,}102 & -- & \textbf{2.6100} $\pm$ 0.0582 & \textbf{0.667\%} & \textbf{1.72\%} & 4.658 \\
\rowhl \textbf{\cappcsa{}-Fast ($q{=}1.5, 4$K)} & Fixed & 5{,}537 & 9{,}492 & -- & 2.7609 $\pm$ 0.0611 & 0.789\% & 2.05\% & 4.733 \\
\rowhl \textbf{\cappcsa{}-Fast ($q{=}1.35, 4$K)} & Fixed & 7{,}385 & 13{,}102 & -- & 2.6256 $\pm$ 0.0586 & \textbf{0.667\%} & \textbf{1.72\%} & 4.658 \\
\rowhl \cappcsa{}-RawEF ($q{=}1.35, 4$K) & Fixed & 7{,}775 & 13{,}102 & -- & 2.7643 $\pm$ 0.0617 & \textbf{0.667\%} & \textbf{1.72\%} & 4.658 \\
Staircase-6b (ours, $8$K) & Fixed & 6{,}880 & 16{,}384 & -- & 3.2392 $\pm$ 0.0747 & 0.767\% & 1.94\% & 7.714 \\
Orig-ELL-28b ($m{=}2048$) & Fixed & 7{,}168 & 8{,}192 & -- & 3.5561 $\pm$ 0.0783 & 0.787\% & 2.06\% & 4.064 \\
Orig-ELL-32b ($m{=}2048$) & Fixed & 8{,}192 & 8{,}192 & -- & 3.6122 $\pm$ 0.0794 & 0.742\% & 1.88\% & \textbf{3.612} \\
ULL-8b ($m{=}8192$) & Fixed & 8{,}192 & 8{,}192 & -- & 4.5150 $\pm$ 0.1028 & 0.830\% & 2.13\% & 4.515 \\
Orig-ELL-32b + \texttt{zstd} & Var & 6{,}412 (max 6{,}482) & 8{,}192 & -- & 2.8272 (var) & 0.742\% & 1.88\% & 3.612 \\
Apache CPC (\texttt{lgK}{=}13, ICON) & Var & 4{,}932 (emp 5{,}060) & 10{,}240 / 75{,}776 & -- & 2.3057 $\pm$ 0.0522 & 0.764\% & 1.97\% & 4.79 / 35.4 \\
\bottomrule
\end{tabular}
\end{table*}

%% file: sections/06_evaluation.tex
\section{Experimental Evaluation}
\label{sec:eval}

\input{tables/tab_latency_tree}
\input{tables/tab_olap_shuffle}

We test \cappcsa{} (\textsc{Exact}, \textsc{Fast}, \textsc{RawEF}) on Apple~M5, Xeon~6430, and Parquet on six workloads: 4k-seed MVP and 16k spills; 0-alloc ops, insert regimes, and 32-shard trees; 16-phase sweeps; shuffles; rollups; and Parquet.

\mypara{Hardware, Toolchain, and Fair Protocol}
{\sloppy Primary runs use Apple~M5 (10c, 24\,GB, macOS 26.6; \texttt{rustc 1.95.0 -O}, \texttt{clang++ 21.0.0 -O3}), checked single-core (\texttt{taskset -c 0}) on Xeon Gold~6430 (x86-64, \texttt{rustc 1.100.0}, C++ \texttt{-march=native}), with DataSketches \texttt{v5.2.0}~\cite{datasketches}, \texttt{libzstd}~v1.5.7~\cite{collet2021zstd}, \texttt{pyarrow}, and \texttt{python-flint 0.6.0}~\cite{johansson2017arb}. Rust baselines share SplitMix64 streams and unified MLE (ULL/ELL also have closed-form estimators~\cite{ertl2024ull,ertl2025ell}); CPC uses MurmurHash3+ICON. Routine coding assistance used generative AI tools, with all proofs, code, and runs verified by the authors (\S\ref{sec:conclusion}). Heap excludes \Nm{fast.static.min.kb}--\Nm{fast.static.max.kb}\,KB LUTs ($0$\,B \textsc{RawEF}/ULL/HLL); Wire MVP is $8 B_{\mathrm{wire}}\mathrm{RelMSE}$.\par}

\subsection{4,000-Seed Showdown and 16k-Trial Spills}
\label{sec:showdown}

Table~\ref{tab:showdown_main} reports 16-phase real-hash sweeps (2k/ph) and 4,000 paired seeds at $n{=}10^6$ (${\le}4$\,KB, ${\le}8$\,KB). Over 16 phases, \textsc{Fast} hits \textbf{\Nm{ph.b16.fast135.mean2}--\Nm{ph.b16.fast15.mean2}} mean MVP (max \Nm{ph.b16.fast135.max}--\Nm{ph.b16.fast15.max}), cutting \Nm{ph.cut.fast15.ell28}\%--\Nm{ph.cut.fast135.ell28}\% vs.\ ELL-28b (\Nm{ph.b16.ell28.mean2}), \Nm{ph.cut.fast15.ell32}\%--\Nm{ph.cut.fast135.ell32}\% vs.\ ELL-32b (\Nm{ph.b16.ell32.mean2}), and \Nm{ph.cut.fast15.curt}\%--\Nm{ph.cut.fast135.curt}\% vs.\ Staircase-6b (\Nm{ph.b16.curt.mean2}). At $n{=}10^6$, \textsc{Fast} ($q{=}1.35$) hits \textbf{\Nm{s1m.fast135.2k.mvp}} in Tier~1 ($-\Nm{cut.fast.2k.ell32.pct}\%$ vs.\ ELL-32b, $z{=}\Nm{pair.fast2k.ell32.z}$) and \textbf{\Nm{s1m.fast135.4k.mvp}} in Tier~2 ($-\Nm{cut.fast.4k.ell32.pct}\%$ vs.\ ELL-32b \Nm{s1m.ell32.2k.mvp}, $-\Nm{s1m.cut.t2.ell28}\%$ vs.\ ELL-28b \Nm{s1m.ell28.2k.mvp}), trading RAM MVP (\Nm{s1m.fast.ram.mvp.lo}--\Nm{s1m.fast.ram.mvp.hi} vs.\ ELL-32b \Nm{s1m.ell32.ram.mvp.lo}--\Nm{s1m.ell32.ram.mvp.hi}) for wire compactness.

\mypara{Variable Framing \& Empirical vs.\ Provable Spills}
ULL-8b+\texttt{zstd} (\Nm{s1m.zstd.ull4k.mvp}) and CPC \texttt{lgK}{=}12 (\Nm{cpc12.mvp}) shrink mean bytes (1\,B raw tag: $4{,}097$\,B, \Nm{s1m.ull.4k.fb.mvp3}). In Table~\ref{tab:parquet_xeon}(B)'s 16k trials, ULL+\texttt{zstd} ($m{=}4{,}096$) hits max \Nm{ullz.16k.maxb}\,B (0/16k spills at $W{=}2{,}400$\,B, eff.\ MVP \Nm{ullz.sp2400.mvp}); CPC \texttt{lgK}{=}12 averages \Nm{cpc12.mean.b}\,B ($n{=}10^6$; ph means \Nm{cpc12.sp16k.ph.lo}--\Nm{cpc12.sp16k.ph.hi}\,B, max \Nm{cpc12.16k.maxb}\,B), hitting 0/16k at \Nm{cpc12.empcap.b}\,B (\Nm{cpc12.sp2580.mvp}) and $\delta{=}\Nm{cpc12.sp2500.delta}\%$ at $2{,}500$\,B (\Nm{cpc12.sp2500.mvp}), vs.\ \cappcsa{}'s \emph{provable} $\delta{=}0$ (\Nm{fast.q15.2k.wire}\,B).

\begin{figure*}[t]
  \centering
  \includegraphics[width=0.53\textwidth]{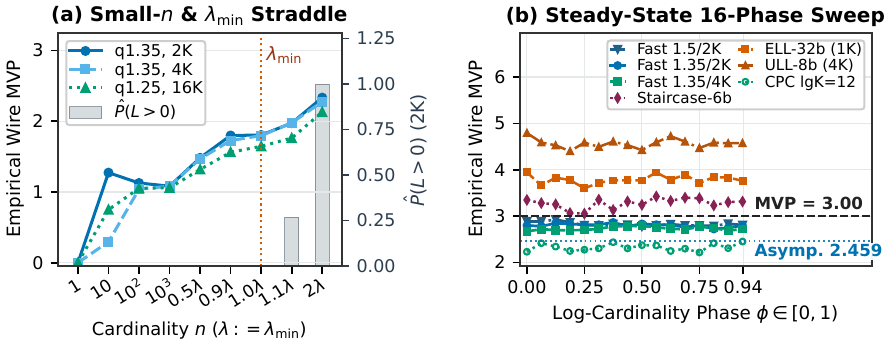}
  \Description{Two-panel plot showing small-cardinality wire MVP and floor-lift probability across lambda_min, and steady-state wire MVP across 16 phases including 16k-trial CapPCSA-Fast, 2k/phase real-hash Staircase-6b, Orig-ELL-32b, ULL-8b, and 16k CPC lgK=12.}
  \caption{\textbf{(a)} Small-$n$ \& $\lambda_{\min}$ (\Nm{small.cells} cells, \Nm{small.trials} trials): $0/4{,}000$ lifts ${\le}0.9\lambda_{\min}$ ($|\text{RelBias}|{\le}\Nm{small.max.bias}\%$, \textsc{Exact} MVP${\le}\Nm{small.max.mvp}$). \textbf{(b)} 16-phase sweep: 16k \textsc{Fast} pointwise 95\% upper CIs stay $<3.00$, alongside 2k/ph Staircase-6b, ELL-32b, ULL-8b, and 16k CPC \texttt{lgK}{=}12.}
  \label{fig:phases_smalln}
\end{figure*}

\input{tables/tab_parquet_xeon_spill}

\subsection{Dual-Host Operators, Regimes, and Trees}
\label{sec:latency}

In Table~\ref{tab:latency_tree} ($n{=}10^6$, M5), \cappcsa{} merges in \textbf{\Nm{ops.cap.mrg.min}--\Nm{ops.cap.mrg.max}\,$\mu$s} (0\,B heap), beating Staircase-6b (\Nm{ops.cap.vs.curt.min}--\Nm{ops.cap.vs.curt.max}$\times$) and CPC (\Nm{ops.cap.vs.cpc.mrg.min}--\Nm{ops.cap.vs.cpc.mrg.max}$\times$). On Xeon (Table~\ref{tab:parquet_xeon}(B)), SIMD inverts raw merges: ULL-8b (\Nm{xeon.ull4k.mrg}\,$\mu$s), ELL-32b (\Nm{xeon.ell32.mrg}\,$\mu$s), and HLL-6b (\Nm{xeon.hll4k.mrg}\,$\mu$s) beat $q{=}1.5$ (\Nm{xeon.cap15.mrg}\,$\mu$s) by \Nm{xeon.mrg.q15.min}--\Nm{xeon.mrg.q15.max}$\times$ ($q{=}1.35$: \Nm{xeon.mrg.q135.ell}--\Nm{xeon.mrg.q135.hll}$\times$), while \cappcsa{} beats Staircase-6b and CPC. Orig-ELL-32b MLE (\Nm{ops.ell32.1k.mle}\,$\mu$s) trails \cappcsa{} (\Nm{ops.cap135.2k.mle}\,$\mu$s) by \Nm{mle.speedup.2k}$\times$. CPC's insert gap (\Nm{ops.cap.vs.cpc.ins}$\times$ M5 vs.\ \Nm{xeon.cpc.ins.ratio}$\times$ Xeon) stems from MurmurHash3 and its \texttt{lgK}{=}12 spike (\Nm{cpc12.ins.ns} vs.\ \Nm{cpc11.ins.ns}/\Nm{cpc13.ins.ns}\,ns). 32-shard trees give \textbf{0\slash\Nm{tree.rust.n} Rust} / \textbf{0\slash\Nm{tree.cpc.n} CPC} mismatches.

{\sloppy\mypara{Insert-Regime Sensitivity}
Under lazy two-stage hashing (\texttt{insert\_\-regime\_\-sweep}, M5, $q{=}1.5, 2$K), \cappcsa{} runs \Nm{reg.n1m.fast15}\,ns/key vs.\ ELL-32b \Nm{reg.n1m.ell32}\,ns (\Nm{reg.ratio.n1m}$\times$) at distinct $n{=}10^6$, and \Nm{reg.dup8.fast15} vs.\ \Nm{reg.dup8.ell32}\,ns (\Nm{reg.ratio.dup8}$\times$) at $8\times$ duplicates; pre-hashed 100k keys (\Nm{reg.ratio.pre}$\times$) and small streams ($n{=}10^2$--$10^4$, up to \Nm{reg.ratio.n100}$\times$; Xeon \Nm{reg.x86.ratio.min}$\times$--\Nm{reg.x86.ratio.max}$\times$) pay hashing at small $L$.\par}

\subsection{16-Phase Stability and Calibration}
\label{sec:phases_smalln}

Fig.~\ref{fig:phases_smalln}: $0/4{,}000$ lifts ${\le}0.9\lambda_{\min}$ ($|\mathrm{RelBias}|{\le}\Nm{small.max.bias}\%$). On 16k/ph multinomial sweeps, all four \textsc{Fast} 95\% upper CIs stay $<3.00$ (\Nm{p16.q135.4k.fast.ci}--\Nm{p16.q15.2k.fast.ci}; Bonferroni: three $<3.00$, $q{=}1.5,2$K spans $3.01$--$3.02$). Over 2k SplitMix64/ph, \textsc{Fast} CIs hit \Nm{ph.b16.fast135.ci2}--\Nm{ph.b16.fast15.ci2} ($g{=}2{,}048$), beating Staircase-6b (\Nm{ph.b16.curt.max}), ELL-32b (\Nm{ph.b16.ell32.max}), and ULL-8b (\Nm{ph.b16.ull.max}); 12-cell 64b vs.\ 192b ELL CIs span $0$ ($[\Nm{elleq.ci.lo}, +\Nm{elleq.ci.hi}]$).

\subsection{OLAP Shuffles, Rollups, and Parquet}
\label{sec:olap}
\label{sec:catalog}

\mypara{OLAP Shuffles (Table~\ref{tab:olap_shuffle}A)}
{\sloppy On sparse Zipf ($P{=}64$, $G{=}512$, \Nm{olap.events} keys), \textsc{Fast} ($q{=}1.5$, $2$K) cuts wire \Nm{olap.cut.15.ell32}\% vs.\ ELL-32b/ULL-8b (\Nm{olap.fast15.mb} vs.\ \Nm{olap.ell32.mb}\,MB; RMSE \Nm{olap.sparse.fast15.rmse}\% vs.\ \Nm{olap.sparse.ell32.rmse}\%/\Nm{olap.sparse.ull.rmse}\%), CPC hits \textbf{\Nm{olap.sparse.cpc12.mb}\,MB}, and raw registers lead $>1$\,Gbps. On dense shuffles ($2{,}048{\times}10^5$ keys), \textsc{RawEF} ($q{=}1.5$) hits \textbf{\Nm{olap.dense.rawef15.ms}\,ms} M5 (\textbf{\Nm{xeon.dense.rawef15.ms}\,ms} Xeon), beating Staircase-6b/CPC/ULL+\texttt{zstd} (\Nm{olap.dense.ullz.ms}\,ms), breaking even at \Nm{olap.dense.rawef15.bw.ull}--\Nm{olap.dense.rawef15.bw.ell32}\,Mb/s/core (\textsc{Fast}: \Nm{olap.dense.bw.ull}--\Nm{olap.dense.bw.ell32}) vs.\ raw.\par}

\mypara{Rollups and Parquet (Tab.~\ref{tab:olap_shuffle}B, \ref{tab:parquet_xeon}A)}
{\sloppy On M5 rollups, \textbf{\textsc{RawEF}} ($q{=}1.5$) runs \textbf{\Nm{cat.r256.rawef15}\,ms} at $R{=}256$ (\Nm{cat.rawef15.vs.cpc.min}--\Nm{cat.rawef15.vs.cpc.max}$\times$ vs.\ CPC, \Nm{cat.rawef15.vs.ullz.min}--\Nm{cat.rawef15.vs.ullz.max}$\times$ vs.\ ULL+\texttt{zstd}; Xeon: ULL+\texttt{zstd} \Nm{xeon.r256.ullz.ms} vs.\ \Nm{xeon.r256.rawef15.ms}\,ms; raw leads RAM). In 4k Parquet (\texttt{Rollup U/S}, excl.\ I/O), \cappcsa{} leads \texttt{FLBA} \texttt{NONE} (\Nm{pq.u.fast15.none}\,B) and \texttt{LZ4} (\Nm{pq.u.fast15.lz4}\,B; skew \texttt{ZSTD} \textbf{\Nm{pq.s.rawef15.zstd}\,B}; \texttt{BYTE\_\allowbreak ARRAY}~\cite{datasketches} CPC/\texttt{ULL+z} store less); on 16-ph \texttt{ZSTD} MVP, \textsc{Fast} (\Nm{pq.u.fast135.zstd.mvp16}--\Nm{pq.u.fast15.zstd.mvp16}), ULL-8b (\Nm{pq.u.ull.zstd.mvp16}), and CPC (\Nm{pq.u.cpc.zstd.mvp16}) tie within 1--3\% (ULL-8b lowest \texttt{ZSTD}, \Nm{pq.u.ull.zstd}\,B).\par}

%% file: tables/tab_latency_tree.tex
\begin{table*}[t]
\centering
\caption{\textbf{\textsc{CapPCSA-Fast} merges in \Nm{ops.cap.mrg.min}--\Nm{ops.cap.mrg.max}\,$\mu$s (0\,B heap) and serializes in \Nm{ops.cap15.2k.fast.enc}--\Nm{ops.cap135.2k.fast.dec}\,$\mu$s ($g{=}2{,}048$), beating CPC \texttt{lgK}{=}12 insert (up to $\Nm{ops.cap.vs.cpc.ins}\times$ M5 / $\Nm{xeon.cpc.ins.ratio}\times$ Xeon) and disjoint merge ($\Nm{ops.cap.vs.cpc.mrg.min}\times$--$\Nm{ops.cap.vs.cpc.mrg.max}\times$), with 0/\Nm{tree.rust.n} Rust and 0/\Nm{tree.cpc.n} CPC mismatches.} M5 unified MLE ($n{=}10^6$); \textsc{RawEF} 32-sh.\ scaled from identical states ($\times 2989/2843$, $3990/3793$, $\pm 10\%$ SE).}
\label{tab:latency_tree}
\footnotesize
\tblclean
\renewcommand{\arraystretch}{1.00}
\setlength{\tabcolsep}{3.1pt}
\begin{tabular}{@{}l r r r r r r r r@{}}
\toprule
\textbf{Sketch} & \textbf{Ins} $\downarrow$ & \textbf{Mrg Disj/Ov50} $\downarrow$ & \textbf{Mrg Heap} $\downarrow$ & \textbf{Enc / Dec} $\downarrow$ & \textbf{Est} $\downarrow$ & \textbf{Disj 32-Sh.} $\downarrow$ & \textbf{8$\times$ Ov 32-Sh.} $\downarrow$ & \textbf{MM} \\
 & (ns) & ($\mu$s) & (B) & ($\mu$s) & ($\mu$s) & Wire MVP ($\pm$SE) & Wire MVP ($\pm$SE) & (/400) \\
\midrule
\rowhl \textbf{\cappcsa{}-Fast ($q{=}1.5, 2$K)} & 8.90 & 3.12 / 3.12 & 0 & 15.9 / 32.5 & 2.39 & 2.620 $\pm$ 0.260 & 2.506 $\pm$ 0.243 & 0/400 \\
\rowhl \cappcsa{}-RawEF ($q{=}1.5, 2$K) & 8.90 & 3.12 / 3.12 & 0 & 5.7 / 6.3 & 2.39 & 2.755 $\pm$ 0.274 & 2.635 $\pm$ 0.255 & 0/400 \\
\rowhl \textbf{\cappcsa{}-Fast ($q{=}1.35, 2$K)} & 12.33 & 3.29 / 3.33 & 0 & 22.5 / 34.3 & 3.20 & 2.879 $\pm$ 0.254 & 2.663 $\pm$ 0.255 & 0/400 \\
\rowhl \cappcsa{}-RawEF ($q{=}1.35, 2$K) & 12.33 & 3.29 / 3.33 & 0 & 9.5 / 10.5 & 3.20 & 3.029 $\pm$ 0.267 & 2.802 $\pm$ 0.269 & 0/400 \\
\rowhl \textbf{\cappcsa{}-Fast ($q{=}1.5, 4$K)} & 13.61 & 2.46 / 2.46 & 0 & 30.0 / 47.7 & \textbf{2.18} & 2.794 $\pm$ 0.248 & 2.551 $\pm$ 0.264 & 0/400 \\
\rowhl \textbf{\cappcsa{}-Fast ($q{=}1.35, 4$K)} & 21.30 & 4.00 / 3.92 & 0 & 42.5 / 65.5 & 3.38 & 2.999 $\pm$ 0.276 & 2.596 $\pm$ 0.283 & 0/400 \\
Staircase-6b (ours, $4$K) & 10.47 & 30.62 / 29.96 & 0 & 32.7 / 18.8 & 43.29 & 3.486 $\pm$ 0.359 & 3.419 $\pm$ 0.389 & 0/400 \\
Orig-ELL-32b ($m{=}1024$) & 10.00 & 3.17 / 3.17 & 0 & \textbf{0.12} / \textbf{0.12} & 27.29 & 4.015 $\pm$ 0.395 & 3.776 $\pm$ 0.371 & 0/400 \\
Orig-ELL-32b ($m{=}2048$) & 10.31 & 6.04 / 6.08 & 0 & 0.23 / 0.24 & 49.67 & 4.054 $\pm$ 0.374 & 3.111 $\pm$ 0.274 & 0/400 \\
ULL-8b ($m{=}4096$) & 6.95 & \textbf{1.25} / \textbf{1.25} & 0 & \textbf{0.12} / \textbf{0.12} & 14.92 & 4.917 $\pm$ 0.478 & 5.306 $\pm$ 0.691 & 0/400 \\
Dense-HLL-6b ($m{=}4096$) & \textbf{4.02} & 2.80 / 2.33 & 0 & 2.9 / 3.5 & 8.08 & 6.595 $\pm$ 0.634 & 6.699 $\pm$ 0.793 & 0/400 \\
ULL-8b ($m{=}4096$) + \texttt{zstd} & 6.95 & 1.25 / 1.25 & 0 & 48.5 / 16.4 & 14.92 & 2.683 (var) & 2.895 (var) & -- \\
Apache CPC (\texttt{lgK}{=}12) & 52.37 & 21.3 / 21.5 & 65{,}536 & 6.1 / 23.2 & 0.007 & 2.260 $\pm$ 0.201 & 2.507 $\pm$ 0.242 & 0/400 \\
Apache CPC (\texttt{lgK}{=}13) & 43.54 & 52.2 / 52.4 & 131{,}072 & 14.2 / 46.7 & 0.007 & 2.115 $\pm$ 0.181 & 2.513 $\pm$ 0.281 & 0/400 \\
\bottomrule
\end{tabular}
\end{table*}

%% file: tables/tab_olap_shuffle.tex
\begin{table*}[t]
\centering
\caption{\textbf{Concurrent-$G$ OLAP: \textsc{RawEF} ($q{=}1.5$) cuts dense CPU to \Nm{olap.dense.rawef15.ms}\,ms (\textbf{61.4\,ms} Xeon) and beats CPC \texttt{lgK}{=}12 on M5 rollups ($\Nm{cat.rawef15.vs.cpc.min}\times$--$\Nm{cat.rawef15.vs.cpc.max}\times$); CPC leads sparse shuffle (\Nm{olap.sparse.cpc12.mb}\,MB).} A: M5 ($P{=}64, G{=}512$ sparse, $2{,}048{\times}10^5$ dense). B: 100k-slot \& 256-part rollups (\textsc{Exact} \Nm{cat.r16.exact.ms}\,ms@$R{=}16$).}
\label{tab:olap_shuffle}
\footnotesize
\tblclean
\renewcommand{\arraystretch}{1.00}
\setlength{\tabcolsep}{3.0pt}
\begin{tabular}{@{}l r r r r r r r r@{}}
\toprule
\multicolumn{9}{@{}l}{\textit{Panel A: Interleaved Concurrent-$G$ OLAP Shuffle (Sparse Zipfian $P{=}64, G{=}512$ + Dense Heavy-Hitter $2{,}048{\times}10^5$ keys)}} \\
\midrule
\textbf{Sketch} & \textbf{Sparse Shuf.} & \textbf{RMSE All/Hvy} & \textbf{Map Ins} & \textbf{Enc / Dec} & \textbf{Red.\ Mrg} & \textbf{Model @1G} & \textbf{Dense Shuf.} & \textbf{Dense Codec+Mrg} \\
 & (MB) & (\%) & (s) & (s) & (s) & (s) & (MB) & (ms) \\
\midrule
\rowhl \textbf{\cappcsa{}-Fast ($q{=}1.5, 2$K)} & 93.16 & 0.775 / 0.862 & 1.609 & 0.763 / 1.194 & 0.1545 & 4.466 & 5.82 & 358.5 \\
\rowhl \textbf{\cappcsa{}-RawEF ($q{=}1.5, 2$K)} & 97.94 & 0.775 / 0.862 & 1.609 & 0.153 / 0.212 & 0.1545 & 2.912 & 6.12 & 64.8 \\
\rowhl \textbf{\cappcsa{}-Fast ($q{=}1.35, 2$K)} & 124.29 & \textbf{0.649} / \textbf{0.713} & 1.958 & 0.926 / 1.510 & 0.2377 & 5.626 & 7.77 & 355.2 \\
\rowhl \cappcsa{}-RawEF ($q{=}1.35, 2$K) & 130.74 & \textbf{0.649} / \textbf{0.713} & 1.958 & 0.214 / 0.324 & 0.2377 & 3.779 & 8.17 & 88.7 \\
\rowhl \cappcsa{}-Exact ($q{=}1.5, 2$K) & \textbf{92.14} & 0.775 / 0.862 & 1.609 & 6.009 / 7.872 & 0.1545 & 16.381 & \textbf{5.76} & 26{,}647.2 \\
Staircase-6b (ours, $4$K) & 112.20 & 0.788 / 0.862 & 2.179 & 1.508 / 0.429 & 0.9975 & 6.011 & 7.01 & 228.4 \\
Orig-ELL-32b ($m{=}1024$) & 134.22 & 0.795 / 0.860 & 0.753 & 0.016 / \textbf{0.006} & 0.1735 & 2.022 & 8.39 & 11.0 \\
ULL-8b ($m{=}4096$) & 134.22 & 0.874 / 0.953 & \textbf{0.632} & \textbf{0.013} / 0.010 & \textbf{0.0713} & \textbf{1.800} & 8.39 & \textbf{3.0} \\
ULL-8b + \texttt{zstd} (Var) & 15.33 & 0.870 / 0.952 & 0.397 & 0.683 / 0.321 & 0.6129 & 2.130 & 4.58 & 188.2 \\
Apache CPC \texttt{lgK}{=}12 (Var) & 9.38 & 0.693 / 0.767 & 1.034 & 0.254 / 0.199 & 0.1297 & 1.700 & 5.11 & 138.5 \\
\midrule
\multicolumn{9}{@{}l}{\rule{0pt}{2.2ex}\textit{Panel B: \S\ref{sec:catalog} Catalog Payload Projection \& In-Memory Rollups ($100{,}000$ projected slots; rollup $R$ of 256 partitions $\to$ merge $\to$ MLE)}} \\
\midrule
\textbf{Catalog Format} & \textbf{Contract} & \textbf{Slot (B)} $\downarrow$ & \textbf{100k Cat.} $\downarrow$ & \textbf{Wire MVP} $\downarrow$ & \textbf{$R{=}16$ P50} $\downarrow$ & \textbf{$R{=}64$ P50} $\downarrow$ & \textbf{$R{=}256$ P50} $\downarrow$ & \textbf{vs.\ CPC12} \\
 &  &  & (MB) & (@$10^6$) & ($\mu$s) & ($\mu$s) & ($\mu$s) & ($R{=}16$ speed) \\
\midrule
\rowhl \textbf{\cappcsa{}-Fast ($q{=}1.5, 2$K)} & Fixed & \textbf{2{,}843} & \textbf{284.3} & \textbf{2.65} & 574.6 & 3{,}160.3 & 13{,}711.8 & 1.03$\times$ \\
\rowhl \textbf{\cappcsa{}-RawEF ($q{=}1.5, 2$K)} & Fixed & 2{,}989 & 298.9 & 2.79 & 146.4 & 611.2 & 3{,}177.1 & 4.05$\times$ \\
\rowhl \textbf{\cappcsa{}-Fast ($q{=}1.35, 2$K)} & Fixed & 3{,}793 & 379.3 & 2.72 & 806.3 & 4{,}070.4 & 19{,}149.0 & 0.74$\times$ \\
\rowhl \cappcsa{}-RawEF ($q{=}1.35, 2$K) & Fixed & 3{,}990 & 399.0 & 2.86 & 235.8 & 1{,}201.8 & 5{,}057.7 & 2.52$\times$ \\
Staircase-6b (ours, $4$K) & Fixed & 3{,}424 & 342.4 & 3.18 & 728.8 & 2{,}646.5 & 9{,}081.2 & 0.81$\times$ \\
Orig-ELL-32b ($m{=}1024$) & Fixed & 4{,}096 & 409.6 & 3.78 & 76.8 & 229.4 & 846.7 & 7.72$\times$ \\
ULL-8b ($m{=}4096$) & Fixed & 4{,}096 & 409.6 & 4.59 & \textbf{48.4} & \textbf{96.3} & \textbf{324.2} & \textbf{12.26$\times$} \\
Orig-ELL-32b + \texttt{zstd} & Var & 3{,}265 (avg) & 326.5 & 3.02 & 190.0 & 906.6 & 3{,}743.5 & 3.12$\times$ \\
ULL-8b ($m{=}4096$) + \texttt{zstd} & Var & 2{,}235 (avg) & 223.5 & 2.51 & 405.0 & 1{,}580.1 & 8{,}306.0 & 1.46$\times$ \\
Apache CPC (\texttt{lgK}{=}12, ICON) & Var & 2{,}490 (avg) & 249.0 & 2.37 & 593.1 & 2{,}262.4 & 11{,}092.9 & 1.00$\times$ \\
\bottomrule
\end{tabular}
\end{table*}

%% file: tables/tab_parquet_xeon_spill.tex
\begin{table*}[t]
\centering
\caption{\textbf{Parquet (4k rows), x86-64 Xeon, and 16k spills.} A: \cappcsa{} leads \texttt{FLBA} \texttt{NONE}/\texttt{LZ4}; 16-ph \texttt{ZSTD} MVP ($8 B_{\mathrm{ZSTD}}\mathrm{RelMSE}_{16}$) ties \textsc{Fast}/ULL-8b/CPC (1--3\%); \texttt{Rollup U/S}: in-mem dec+mrg. B: Xeon ($R{=}256$ \texttt{ULL+z}/\textsc{RawEF}); right: 0/16k* vs.\ $\delta{=}0$.}
\label{tab:parquet_xeon}
\footnotesize
\tblclean
\renewcommand{\arraystretch}{1.00}
\setlength{\tabcolsep}{2.9pt}
\begin{tabular}{@{}l l r r r r r r r@{}}
\toprule
\multicolumn{9}{@{}l}{\textit{Panel A: Real Apache Parquet Storage (B/row, col.) \& $R{=}256$ Rollup (4,096 rows, rg=1,024; \texttt{NONE}/\texttt{LZ4}/\texttt{ZSTD})}} \\
\midrule
\textbf{Catalog Column} & \textbf{Parquet Type} & \textbf{Unif NONE} & \textbf{Unif LZ4} & \textbf{Unif ZSTD} & \textbf{ZSTD MVP} & \textbf{Skew LZ4} & \textbf{Skew ZSTD} & \textbf{Rollup U/S} \\
 &  & (B/row) & (B/row) & (B/row) & (16-ph) $\downarrow$ & (B/row) & (B/row) & ($R{=}256$, ms) \\
\midrule
\rowhl \textbf{\cappcsa{}-Fast ($q{=}1.5, 2$K)} & \texttt{FLBA} & \textbf{2{,}843.0} & \textbf{2{,}407.2} & 2{,}291.7 & 2.28 & 1{,}754.0 & 1{,}666.9 & 12.07/9.90 \\
\rowhl \textbf{\cappcsa{}-RawEF ($q{=}1.5, 2$K)} & \texttt{FLBA} & 2{,}989.0 & 2{,}670.8 & 2{,}453.7 & 2.44 & \textbf{1{,}683.3} & \textbf{1{,}473.4} & 2.91/2.18 \\
\rowhl \cappcsa{}-Fast ($q{=}1.35, 2$K) & \texttt{FLBA} & 3{,}793.0 & 3{,}228.7 & 3{,}071.5 & 2.26 & 2{,}235.1 & 2{,}121.9 & 16.38/12.28 \\
Staircase-6b (ours, $4$K) & \texttt{FLBA} & 3{,}424.0 & 3{,}322.1 & 3{,}143.3 & 3.01 & 2{,}246.7 & 2{,}092.2 & 11.02/8.99 \\
Orig-ELL-32b ($m{=}1024$) & \texttt{FLBA} & 4{,}096.0 & 4{,}098.3 & 3{,}144.4 & 2.91 & 2{,}580.6 & 2{,}049.8 & 0.78/0.90 \\
ULL-8b Raw ($m{=}4096$) & \texttt{FLBA} & 4{,}096.0 & 4{,}021.4 & \textbf{2{,}066.2} & 2.30 & 2{,}522.0 & 1{,}570.9 & \textbf{0.35}/\textbf{0.35} \\
Apache CPC \texttt{lgK}{=}12 (Var) & \texttt{BYTE\_ARR} & 2{,}493.3 & 2{,}498.8 & 2{,}493.4 & 2.33 & 1{,}426.0 & 1{,}425.8 & 10.70/8.98 \\
ULL-8b + \texttt{zstd} (Var) & \texttt{BYTE\_ARR} & 2{,}238.0 & 2{,}245.1 & 2{,}238.1 & 2.46 & 1{,}391.1 & 1{,}388.7 & 7.37/6.26 \\
\midrule
\multicolumn{6}{@{}l|}{\rule{0pt}{2.2ex}\textit{Panel B (Left): x86-64 Xeon (1 Core, Gold 6430)}} & \multicolumn{3}{l@{} }{\textit{Panel B (Right): CPC/ULL+z Spill (16k)}} \\
\midrule
\textbf{Xeon Gold 6430} & \textbf{Ins} & \textbf{Mrg} & \textbf{Enc/Dec} & \textbf{Dense} & \multicolumn{1}{r|}{\textbf{$R{=}256$}} & \textbf{Slot $W$} & \textbf{Overflow $\delta$} & \textbf{Eff.\ MVP} \\
\midrule
\rowhl \cappcsa{}-Fast ($q{=}1.5, 2$K) & 8.49\,ns & 2.49\,$\mu$s & 19.3/32.0 & 182.0\,ms & \multicolumn{1}{r|}{13.00\,ms} & \texttt{CPC} $2{,}400$\,B & 100.0\% & 4.58 \\
\rowhl \cappcsa{}-RawEF ($q{=}1.5, 2$K) & 8.49\,ns & 2.49\,$\mu$s & 6.2/7.5 & 61.4\,ms & \multicolumn{1}{r|}{3.26\,ms} & \texttt{CPC} $2{,}500$\,B & 33.3\% & 3.12 \\
Orig-ELL-32b ($m{=}1024$) & 7.45\,ns & 0.75\,$\mu$s & 0.09/0.05 & 7.3\,ms & \multicolumn{1}{r|}{0.22\,ms} & \texttt{CPC} $2{,}560$\,B & 0.062\% & 2.39 \\
ULL-8b ($m{=}4096$) [Raw/+z] & 5.32\,ns & 0.54\,$\mu$s & 0.05/0.05 & 7.1/170\,ms & \multicolumn{1}{r|}{0.19/2.75} & \texttt{CPC} $2{,}580$\,B & 0/16k* & 2.41* \\
Apache CPC (\texttt{lgK}{=}12) & 19.5\,ns & 31.4\,$\mu$s & 11.1/22.5 & 158.0\,ms & \multicolumn{1}{r|}{9.05\,ms} & \texttt{ULL+z} $2{,}400$\,B & 0/16k* & 2.64* \\
\bottomrule
\end{tabular}
\end{table*}

%% file: sections/07_related_conclusion.tex
\section{Related Work}
\label{sec:related}

\mypara{Compressed Bitmaps}
{\sloppy PCSA~\cite{flajolet1985pcsa}, MRB~\cite{estan2003bitmap}, arithmetic PCSA~\cite{scheuermann2007compression}, and CPC~\cite{lang2017cpc,datasketches} compress bitmaps. Pettie--Wang~\cite[Thm.~6]{pettie2021fish} proved $H_0/I_0 \approx \Nm{c.pw21.5}$ and Fishmonger~\cite[\S6]{pettie2021fish}; HLLL~\cite{karppa2022hlll} and Huffman-Bucket~\cite{karppa2026hbs} allow $\delta{>}0$. \cappcsa{} pairs caps with~\cite{cover1973enumerative} or rANS to match \Nm{c.merge.5} (\Nm{mw.lo.3}--\Nm{mw.hi.3}, Thm.~\ref{thm:lb}/Cor.~\ref{cor:top_collapse}).\par}

\mypara{In-Place, Anchored \& Non-Mergeable}
{\sloppy LogLog~\cite{durand2003loglog}, HLL/++~\cite{flajolet2007hll,heule2013hllpp}, EHLL~\cite{ohayon2021ehll}, SetSketch~\cite{ertl2021setsketch}, ULL/ELL~\cite{ertl2024ull,ertl2025ell}, and SpikeSketch~\cite{du2023spikesketch,ertl2017hll,wang2023gra} use independent registers; HLL-TailCut~\cite{xiao2020tailcut} and \texttt{HLL\_4}~\cite{datasketches} share a base (3b~\cite{xiao2020tailcut}; \texttt{HLL\_4} exceptions, $\delta{>}0$). Curtain~\cite{pettie2021nonmergeable} anchors (\Nm{curtain.3b.mvp} vs.\ Staircase-6b, Rem.~\ref{rem:anchored}); Thm.~\ref{thm:ell_div} proves ELL divergence $(\ln\Lambda{+}\ln\ln\Lambda)/I_0{+}O(1)$. Martingales~\cite{cohen2014hip,ting2014streamed,pettie2021nonmergeable} hit $H_0/2{=}\Nm{c.pwy21.5}$ (rate \Nm{c.rate.6}, Thm.~\ref{thm:lb}(b), Rem.~\ref{rem:scope}); \cite{janson2025hypertwobits} merges approx.; Dynamic\-Log\-Log~\cite{bushnell2026dll}/AVLL~\cite{bushnell2026avll} overflow asynchronously (Table~\ref{tab:applicability}). Streaming bounds~\cite{kane2010optimal,blasiok2018optimal,karayel2023optimal} pin $O(\varepsilon^{-2}{+}\log n)$~\cite{wang2025thesis}.\par}

\section{Conclusion and Scope}
\label{sec:conclusion}

We prove $\delta{=}0$ limits \Nm{c.merge.6} (64b \textbf{\Nm{mw.lo.3}--\Nm{mw.hi.3}}, factor \Nm{d64.mrg.arb.3}) and \Nm{c.rate.6}\slash\Nm{d64.rate.arb.3}, cover $\top$-collapses (Cor.~\ref{cor:top_collapse}), and prove ELL divergence $(\ln\Lambda{+}\ln\ln\Lambda)/I_0{+}O(1)$ (Thm.~\ref{thm:ell_div}). \cappcsa{} matches both (UB $<3.00$, 5 configs; \textsc{RawEF} $\delta{=}0$): \textsc{Fast} leads fixed-wire/\texttt{FLBA} \texttt{NONE}/\texttt{LZ4} (tying ULL-8b/CPC on 16-ph \texttt{ZSTD} MVP), \textsc{RawEF} leads CPC on M5 rollups/dense OLAP, raw registers lead x86-64 merges/RAM, and CPC leads sparse shuffles (\S\ref{sec:beyond_lin} non-linearizable open).

\mypara{AI-Assistance Disclosure}
Generative AI tools assisted with language polishing, formatting, and routine coding; the authors independently verified all proofs, code, and experiments and bear full responsibility.

%% file: sections/appendix_merged.tex
\section{Supplemental Proof Details for Lemma~2 and Theorem~1}

For reference, the atomic kernels $\dot{H}(t) = h_2(1{-}e^{-t})$, $\dot{I}(t) = t^2/(e^t - 1)$, and $\dot{R}(t) = 2te^{-t}$ satisfy
\begin{align*}
  H_0 &= \int_0^\infty \dot{H}(u)\,\frac{du}{u} = \Nm{h0.6}\ldots, \\
  I_0 &= \int_0^\infty \dot{I}(u)\,\frac{du}{u} = \frac{\pi^2}{6} = \Nm{i0.6}\ldots, \\
  \int_0^\infty \dot{R}(u)\,\frac{du}{u} &= 2.
\end{align*}
Let $H_{\mathrm{upper}} := \int_0^{\ln 2} \dot{H}(u)\,\frac{du}{u} = \Nm{hupper.6}\ldots$ and define $G(t) := \int_0^t \widetilde{H}(u)\,\frac{du}{u}$. For $t > \ln 2$, $G(t) = H_{\mathrm{upper}} + \ln(t/\ln 2)$ strictly; extending that logarithmic branch below $\ln 2$ introduces spurious roots at $t \approx \Nm{root.spur.merge}$ ($\dot{I}$) and $t \approx \Nm{root.spur.rate}$ ($\dot{R}$).

\mypara{Exact Analytic Series for $G(t)$ on $(0, \ln 2]$ (360-Bit Arb)}
Substituting $v = 1 - e^{-u} \in (0, 1/2]$ ($du/u = dv / ((1{-}v)(-\ln(1{-}v)))$) gives
\[
  G(t) = \frac{1}{\ln 2}\!\left[p + \int_0^p (-\ln v)\,\phi(v)\,dv\right], \quad p := 1 - e^{-t},
\]
where $\phi(v) := \frac{v}{(1-v)(-\ln(1-v))} = \frac{1}{1 - \sum_{k=1}^\infty \frac{v^k}{k(k+1)}} = \sum_{m=0}^\infty c_m v^m$ with $c_0 = 1$ and $c_m = \sum_{k=1}^m \frac{c_{m-k}}{k(k+1)} \in (0, 1]$. Term-by-term integration gives $\int_0^p (-\ln v)v^m\,dv = p^{m+1}\bigl(\frac{-\ln p}{m+1} + \frac{1}{(m+1)^2}\bigr)$, with positive tail beyond $m=N$ strictly bounded by $\frac{p^{N+2}}{1-p}\bigl(\frac{-\ln p}{N+2} + \frac{1}{(N+2)^2}\bigr) < 2^{-360}$ at $N{=}360$, yielding a zero-quadrature 360-bit FLINT/Arb enclosure matching 55-digit \texttt{mpmath} to $< 10^{-35}$.

\mypara{Strict Monotonicity of $r_K(t) = \widetilde{H}(t)/K(t)$ on $(0, \ln 2]$}
Since $\widetilde{H}(t) = \dot{H}(t)$ on $(0, \ln 2]$ and $-\ln(1 - e^{-t}) = \sum_{n=1}^{\infty} \frac{e^{-nt}}{n}$, we have $-(1 - e^{-t})\ln(1 - e^{-t}) = e^{-t}\sum_{n=1}^{\infty}\frac{1 - e^{-nt}}{n(n+1)}$, so
\[
  r_{\dot{R}}(t) = \frac{\dot{H}(t)}{\dot{R}(t)} = \frac{1}{2\ln 2}\!\left[1 + \sum_{n=1}^{\infty} \frac{1 - e^{-nt}}{n(n+1)\,t}\right].
\]
Each summand $(1 - e^{-nt})/t = \frac{1}{t}\int_0^t n e^{-nu}\,du$ is the running average of a strictly decreasing positive function, so $r_{\dot{R}}$ strictly decreases on $(0,\ln 2]$. Likewise $\dot{R}(t)/\dot{I}(t) = 2(1 - e^{-t})/t$ strictly decreases, so $r_{\dot{I}} = r_{\dot{R}} \cdot (\dot{R}/\dot{I})$ strictly decreases on $(0,\ln 2]$, proving $F_K'(t) < 0$ on $(0, \ln 2]$.

\mypara{Universal Analytic Proof of $\Delta(\alpha) \le 0$ on $[\alpha_1(\tau), \alpha_0]$ and 360-Bit Arb $c_K(\tau)$}
Let $\Delta(\alpha) := G(\alpha e^{-\tau}) - C J_K(\alpha)$ on $(0, \alpha_0]$ ($\alpha_0 = e^\tau \ln 2$, $\Delta(\alpha_0) \le 0$ by condition~(ii)). By the series above,
\begin{align*}
  \Psi(u) := \frac{\dot{H}(u)}{u} &= \frac{e^{-u}}{\ln 2}\!\left(1 + \sum_{n=1}^{\infty}\frac{1}{n(n+1)}\!\int_0^n e^{-tu}\,dt\right) \\
  &= \int_{[1,\infty)} e^{-su}\,d\nu(s), \qquad d\nu(s) := \frac{\delta_1 + \lfloor s\rfloor^{-1}\,ds}{\ln 2}.
\end{align*}
Since $\nu > 0$ is not a single point mass, Cauchy--Schwarz gives $\Psi\Psi'' - (\Psi')^2 > 0$, so $\ln(\dot{H}(u)/u)$ is strictly convex on $(0,\infty)$. Setting $E_f(x) := x f'(x)/f(x)$ yields $\frac{d}{du}\ln(\dot{H}(u)/u) = (E_{\dot{H}}(u){-}1)/u$, hence $u E_{\dot{H}}'(u) + 1 - E_{\dot{H}}(u) > 0$ for all $u > 0$. Meanwhile, $E_{\dot{R}}(\alpha) = 1{-}\alpha$ gives $-\alpha E_{\dot{R}}'(\alpha) = 1{-}E_{\dot{R}}(\alpha)$, and $E_{\dot{I}}(\alpha) = 2 - \alpha/(1{-}e^{-\alpha})$ satisfies $-\alpha E_{\dot{I}}'(\alpha) - (1{-}E_{\dot{I}}(\alpha)) = 4e^{-\alpha}(\sinh^2(\alpha/2) - (\alpha/2)^2)/(1{-}e^{-\alpha})^2 > 0$ ($\alpha > 0$). Thus at any zero of $g_\tau(\alpha) := E_{\dot{H}}(\alpha e^{-\tau}) - E_K(\alpha)$ on $(0,\alpha_0]$, $\alpha g_\tau'(\alpha) = u E_{\dot{H}}'(u) - \alpha E_K'(\alpha) \ge u E_{\dot{H}}'(u) + 1 - E_{\dot{H}}(u) > 0$ ($u := \alpha e^{-\tau}$), so $g_\tau$ changes sign at most once ($-\!\to\!+$). Because $\frac{d}{d\alpha}\ln(\dot{H}(\alpha e^{-\tau})/K(\alpha)) = g_\tau(\alpha)/\alpha$ and $\dot{H}(\alpha e^{-\tau})/K(\alpha) \to +\infty$ as $\alpha\to 0^+$, $\Delta'(\alpha) = \frac{K(\alpha)}{\alpha}(\dot{H}(\alpha e^{-\tau})/K(\alpha) - C)$ changes sign at most twice on $(0,\alpha_0]$ in pattern $(+)\!\to\!(-)$ or $(+)\!\to\!(-)\!\to\!(+)$. With $\Delta(0^+){=}0$, $\Delta > 0$ near $0^+$, and $\Delta(\alpha_0) \le 0$, beyond its first positive root $\alpha_1(\tau) \in (0,\alpha_0]$ the function $\Delta$ stays $\le 0$ on all of $[\alpha_1(\tau), \alpha_0]$ for every admissible $\tau > 0$ (independently checked on $1{,}500$ pinned 360-bit Arb intervals at $\tau{=}1.80, 1.40$ and $46$ grid shifts).

To certify Eq.~(9) in 360-bit Arb, set $t_0{=}0.05$ and $\alpha_0{=}e^\tau\ln 2$. For $0 < \alpha < t \le t_0$, $(1{-}t_0/2)(-\ln t_0)/\ln 2$ exceeds $c\cdot\sup_{u>0} K(u)/u$; for $\alpha_0 \le \alpha < t$, check $\tau > c(J(\infty) - J_K(\alpha_0))$. On the remaining domain $\{0 < \alpha \le \alpha_0,\; t \ge t_0\}$, separate adaptive 360-bit Arb 1D interval bounds verify $\inf_{t \ge t_0}(G(t) - c J_K(t)) \ge \sup_{0 < \alpha \le \alpha_0}(G(\alpha e^{-\tau}) - c J_K(\alpha))$, certifying $c_{\dot{I}}(1.20) \ge \Nm{d64.cI.arb.5}$ and $c_{\dot{R}}(0.85) \ge \Nm{d64.cR.arb.5}$.

\section{Supplemental Asymptotic Bounds for Theorem~2(5)}

Along the sequence $q_j = 1 + 1/j$, $g_j = 2^{\lceil 6\log_2 j\rceil} \in [j^6, 2j^6)$, $U_j = \lceil e^{j^2}\rceil$, $K_j = \lceil \log_{q_j} U_j\rceil = \Theta(j^3)$, and $h_j = \ln q_j = \frac{1}{j} + O(j^{-2})$:
\begin{enumerate}[leftmargin=*, topsep=2pt, itemsep=2pt]
  \item \textbf{Trigger failure bound.} With $c_m = \lceil g_j(p_m + h_j^2)\rceil$ for $m \in \{s,\dots,m_0{-}1\}$ and $c_{\mathrm{end}} = c_{m_0-1}$, every index $k \in \{m_0,\dots,K_j{-}1\}$ has $p_k \le p_{m_0-1}$, so $c_{\mathrm{end}} - g_j p_k \ge g_j h_j^2$. Hoeffding's inequality bounds every binomial tail in both sums of Eq.~(15) by $\exp(-2g_j h_j^4) = \exp(-\Omega(j^2))$. For the far tail, $\mu_j = g_j T q_j^{-m_0}/(1 - 1/q_j) \le g_j h_j$ by definition of $m_0$, while $C_{\mathrm{tail}} = \lceil 2 g_j h_j\rceil \ge 2\mu_j$, so Poisson Chernoff gives $\Pr[\mathrm{Poi}(\mu_j) > C_{\mathrm{tail}}] \le \exp(-\Omega(g_j h_j)) = \exp(-\Omega(j^5))$. Summing over $K_j = \Theta(j^3)$ levels keeps $\delta_{\mathrm{cap}} = \exp(-\Omega(j^2)) \to 0$.
  \item \textbf{Mixed-radix bit budget.} In Eq.~(14), the $(K_j{+}1)$ floor header adds $\frac{h_j}{g_j}\log_2(K_j{+}1) = O(j^{-7}\log j) = o(1)$. Because $\sum_{u=0}^{c_m} \binom{g_j}{u} \allowbreak \le 2^{g_j h_2(\min(c_m/g_j, 1/2))}$ holds unconditionally, each layer $m \in \{s,\dots,m_0{-}1\}$ adds at most $g_j h_2(\min(c_m/g_j,\allowbreak 1/2))$ bits. For any layer with $c_m/g_j \le 1/2$, we have $h_2(c_m/g_j) \le \widetilde{H}(T q_j^{-m}) + O(h_j^2\ln(1/h_j))$; for the $O(1)$ boundary layers with $c_m/g_j > 1/2$, $p_m > 1/2 - O(h_j^2)$ implies $1 - \widetilde{H}(T q_j^{-m}) = O(h_j^4)$. Since $\widetilde{H}(T q_j^{-m}) = 1$ for $0 \le m < s$, multiplying by $h_j/g_j$ gives the Riemann upper sum $h_j\sum_{m=0}^{m_0-1}\widetilde{H}(T e^{-m h_j}) + O(h_j^2\ln^2(1/h_j)) \le G(T) + o(1)$ (finite-$m_0$ truncation drops nonnegative terms). Using $C_{\mathrm{tail}} \le g_j(K_j{-}m_0)/2$, the far-tail combinadic term is at most $C_{\mathrm{tail}}\log_2(e g_j K_j / C_{\mathrm{tail}}) \allowbreak + O(1)$, contributing $O(h_j^2 \log(K_j/h_j)) = O(j^{-2}\log j) = o(1)$.
  \item \textbf{Uniform phase quadrature.} On $v \in [0,\infty)$, $f_K(v) := K(T e^{-v})$ is smooth, nonnegative, and unimodal with $f_K(\infty)=0$, so its total variation $V_K < \infty$ is finite. Across $J_{\max}(j){+}1$ consecutive levels ($J_{\max}(j) = \min(K_j, \lfloor j^2\rfloor)$) from phase $r \in [0,1]$ with mesh width $h_j$, the Riemann sum $\sum_{m=0}^{J_{\max}(j)} h_j f_K((m{+}r)h_j)$ approximates $\int_0^{(J_{\max}(j)+1)h_j} f_K(v)\,dv$ within $\pm h_j V_K = O(1/j)$ uniformly in $r$, while the omitted tail beyond $(J_{\max}(j){+}1)\allowbreak h_j = \Omega(j)$ vanishes as $e^{-\Omega(j)}$ and the finite-$K_j$ cutoff term in Eq.~(16) contributes at most $c_K h_j (U_j/4)\allowbreak q_j^{-K_j}/g_j \le c_K h_j/(4g_j) = O(j^{-7}) = o(1)$.
\end{enumerate}

\section{Anchored Staircase Closures (Details for Remark~2)}
\setcounter{proposition}{0}
\renewcommand{\theproposition}{C.\arabic{proposition}}

Consider $g$ buckets over levels $1,\dots,K$ with $k$-th largest anchor $A(X) = M_{(k)}(X)$ ($1 \le k \le g$, setting $u_r = 0$ for empty buckets), floor $L(X) = \max(1, A(X) - R + 1)$, interior history depths $d(0),\dots,d(R{-}1)$ ($0 \le d(\Delta) \le R{-}1{-}\Delta$), and overflow depth $d_{\mathrm{ov}} \ge 0$. For each bucket $r$ with maximum hit $u_r$: if $u_r < L$, levels $1\dots L{-}1$ are filled; if $u_r \ge L$, levels $1\dots \ell(A, u_r){-}1$ are filled, where $\ell(A, u_r) := \max(L, u_r - d(A - u_r))$ for $u_r \le A$ and $\max(L, u_r - d_{\mathrm{ov}})$ for $u_r > A$.

\begin{proposition}[Anchored Staircase Homomorphism Conditions]
Consider conditions:
\begin{align*}
(\mathrm{S1})\; &0 \le d(\Delta{-}1) - d(\Delta) \le 1\;\;(1 \le \Delta \le R{-}1),\qquad
(\mathrm{S2})\; d_{\mathrm{ov}} \ge d(0),\\
(\mathrm{S3})\; &d(0) < R{-}1 \Rightarrow d_{\mathrm{ov}} \le d(0)+1.
\end{align*}
(1) If $(\mathrm{S1})$--$(\mathrm{S3})$ hold, $\Phi$ is an exact linearizable join-semilattice closure for all $g, k, R, K$.
(2) Whenever $2 \le k \le g{-}1$ and $K \ge R{+}1$, $(\mathrm{S1})$--$(\mathrm{S3})$ are also necessary for union homomorphism.
\end{proposition}

\begin{proof}[Proof Sketch and Exhaustive Verification]
For sufficiency, $(\mathrm{S1})$ and $(\mathrm{S3})$ ensure $\ell(A, u_r)$ is non-decreasing in $u_r$ at fixed $A$, while $(\mathrm{S1})$ and $(\mathrm{S2})$ ensure $\ell(A, u_r)$ and $L(A)$ are non-decreasing in $A$ at fixed $u_r$. Every surviving bucket ($u_r \ge L$) preserves $u_r$, and every truncated bucket ($u_r < L$) fills strictly below $\max(1, A)$, keeping $A(\Phi(X)) = A(X)$ and $\Phi(\Phi(X)) = \Phi(X)$. Moreover, forcing any cell $(r, j)$ inspects only whether $A \ge j + R$ or ($u_r \ge j + 1$ and $j < \ell(A, u_r)$), both depending exclusively on hits at levels strictly above $j$, so descending level order $K,\dots,1$ topologically linearizes $D(\Phi, \mathcal{C})$. For necessity under any $2 \le k \le g{-}1$ and $K \ge R{+}1$, let $r_1,\dots,r_k$ be $k$ buckets disjoint from a target bucket $r_0$ (all other buckets at $u{=}0$):
(i) \emph{Failure of $d(\Delta{-}1) \ge d(\Delta)$:} set $u_{r_0} = R{-}\Delta{+}1$ and $r_1,\dots,r_k$ to $R$ ($A{=}R, L{=}1$), then lift $r_1,\dots,r_k$ to $R{+}1$ ($A'{=}R{+}1, L'{=}2$): $\Phi(X)$ fills level $u_{r_0} - d(\Delta) \ge 2$ while $\Phi(X')$ leaves it unforced.
(ii) \emph{Failure of $d(\Delta{-}1) - d(\Delta) \le 1$:} keep $r_1,\dots,r_k$ at $R$ and advance $r_0$ from $R{-}\Delta$ to $R{-}\Delta{+}1$: $\Phi(X)$ fills $j^* := R{-}\Delta{-}d(\Delta){-}1 \ge 1$ whereas $\Phi(X')$ leaves $j^*$ unforced.
(iii) \emph{Failure of $(\mathrm{S2})$ or $(\mathrm{S3})$:} for $(\mathrm{S2})$ ($d_{\mathrm{ov}} < d(0)$), put $u_{r_0} = R{+}1$, $r_1,\dots,r_{k-1}$ at $R$, $r_k$ at $0$, then lift $r_1,\dots,r_k$ to $R{+}1$, leaving $\max(2, R{+}1{-}d(0))$ filled in $\Phi(X)$ but unforced in $\Phi(X')$. For $(\mathrm{S3})$ ($d(0) < R{-}1, d_{\mathrm{ov}} \ge d(0)+2$), put $r_1,\dots,r_{k-1}$ at $R$ and advance $r_0$ from $R$ to $R{+}1$ (keeping $A{=}R$ since $k \ge 2$), breaking $\Phi(X) \subseteq \Phi(X')$. Exhaustive checks yield $0$ mismatches across $30$ schedules / $\Nm{thm3.trans}$ transitions ($R{=}3, K{=}5$) and $8$ schedules / $2{,}097{,}152$ pairs ($R{=}2, K{=}3$).
\end{proof}

\section{Exact 64-Bit Parameter Arrays, Lattice Certificates, and Artifact Guide}

\mypara{Exact 64-Bit \cappcsa{} Parameter Tuples (Table~2)}
\input{tables/app_params_list}

\mypara{Small-Lattice Audit Details ($2 \le |\mathcal{S}| \le 8$, Section~3.4)}
Across all \Nm{lat.total} unlabeled finite lattices ($2 \le |\mathcal{S}| \le 8$, \Nm{lat.lin} linearizable vs.\ \Nm{lat.nonlin} non-linearizable), \texttt{scipy} heuristic weight search followed by 360-bit Arb branch-and-bound evaluation of each candidate's fixed-weight objective at $B{=}3$ ($|\mathcal{S}|{=}8$) yields narrow enclosures for the search-selected linearizable vs.\ non-linearizable pair at each tested window (\texttt{ID~159} at $D{=}2$, outward UB \Nm{lat.id159.ub}; \texttt{ID~166} at $D{=}4$, outward UB \Nm{lat.id166.ub}; best search-selected non-linearizable candidates enclose at $2.105446$ for $D{=}1$ and $2.380882$ for $D{=}2$), with $+\Nm{lat.id157.pen}\%$ ($D{=}2$), $+\Nm{lat.id165.pen}\%$ ($D{=}4$), $+\Nm{lat.id218.pen}\%$ ($D{=}8$), and $+\Nm{lat.id212.pen}\%$ ($D{=}64\ln 2$) separation (no counterexample found by heuristic weight search).

\mypara{Artifact Reproduction Guide}
{\sloppy The verification and benchmark scripts below summarize our reproducibility artifact, which will be publicly released upon publication. Primary runs use Apple~M5 (10 cores, 24\,GB RAM, macOS 26.6 / Darwin 25.6 ARM64; \texttt{rustc 1.95.0} \texttt{opt-level=3}, Apple \texttt{clang++ 21.0.0 -O3}), cross-validated single-core (\texttt{taskset -c 0}) on Intel Xeon Gold~6430 (x86-64 \texttt{opt-level=3}, 128 CPUs, \texttt{rustc 1.100.0}; C++ \texttt{-march=native}), with DataSketches C++ \texttt{v5.2.0}, \texttt{libzstd v1.5.7}, \texttt{pyarrow}, and \texttt{python-flint 0.6.0}:\par}
\begin{itemize}[leftmargin=*, topsep=0.5pt, itemsep=0pt, parsep=0pt]\raggedright\footnotesize
  \item \texttt{p1\_constants\_and\_bounds\_v14.py}, \texttt{p2\_cappcsa\_v14\_arb360.py}: 360-bit Arb Theorem~1 bounds (non-vacuous matched \Nm{mw.lo.3}--\Nm{mw.hi.3}; width factor \Nm{d64.mrg.arb.3}/\Nm{d64.mrg.elem.3}) and \textsc{Fast}/\textsc{RawEF} $\delta{=}0$ certs (plus full chain \texttt{scripts/p1\_v12*--p5*}, \texttt{requirements.txt}).
  \item \texttt{rust\_cappcsa/}: zero-dep crate (\texttt{cargo test -{}-release -{}- -{}-test-threads=1}, 16 tests, 0 warnings) for \textsc{Exact}/\textsc{Fast}/\textsc{RawEF}, Lazy 128-bit hash, lazy Staircase-6b, 0-alloc Orig-ELL MLE, ULL, Dense-HLL, \texttt{GlobalAlloc} tests, 2k+ fuzz frames.
  \item \texttt{cpp\_cpc\_bench/cpc\_and\_zstd\_bench\_v14.cpp}, \texttt{scripts/parquet\_catalog\_bench.py}: C++17 CPC~v5.2.0/\texttt{libzstd} harness (showdown, merges, 32-shard trees, OLAP, rollups, 16k spills) and Apache Parquet \texttt{FLBA}/\texttt{BYTE\_ARRAY} bench.
  \item \texttt{verify\_cappcsa\_v14.py}: offline deterministic verifier (Arb-360 bounds, byte budgets, Xeon/Parquet outputs, Rust/C++ trees, SHA-256).
\end{itemize}

%% file: tables/app_params_list.tex
Configs with $g \le 4{,}096$ use $T = 1.25\,t_* \approx 4.197434$; \texttt{q1p25\_g16384} uses $T = 1.18\,t_* \approx 3.962378$. Exact caps $c_{s..m_0-1}$ and $(K, J_{\max}, C_{\mathrm{tail}})$ (4,096 Arb phase subintervals):
\begin{itemize}[leftmargin=*, topsep=1.5pt, itemsep=1pt]\footnotesize\sloppy
  \item \texttt{q2p0\_g2048} ($K{=}64, J_{\max}{=}39, C_{\mathrm{tail}}{=}34$): $c_{3..9} =$ [918, 543, 308, 172, 97, 56, 34].
  \item \texttt{q1p5\_g2048} ($K{=}110, J_{\max}{=}64, C_{\mathrm{tail}}{=}35$): $c_{5..17} =$ [955, 712, 518, 373, 266, 190, 136, 97, 71, 52, 38, 29, 22].
  \item \texttt{q1p35\_g2048} ($K{=}148, J_{\max}{=}84, C_{\mathrm{tail}}{=}37$): $c_{7..24} =$ [909, 731, 580, 456, 357, 278, 216, 169, 132, 103, 81, 64, 51, 41, 33, 27, 22, 18].
  \item \texttt{q1p5\_g4096} ($K{=}110, J_{\max}{=}64, C_{\mathrm{tail}}{=}59$): $c_{5..17} =$ [1860, 1376, 994, 707, 500, 352, 248, 175, 125, 89, 65, 48, 35].
  \item \texttt{q1p35\_g4096} ($K{=}148, J_{\max}{=}84, C_{\mathrm{tail}}{=}62$): $c_{7..24} =$ [1767, 1412, 1114, 870, 675, 522, 403, 311, 240, 186, 144, 113, 88, 70, 55, 44, 35, 29].
  \item \texttt{q1p25\_g16384} ($K{=}199, J_{\max}{=}112, C_{\mathrm{tail}}{=}180$): $c_{8..34} =$ [8210, 7009, 5920, 4958, 4123, 3410, 2808, 2305, 1887, 1543, 1260, 1028, 839, 685, 559, 457, 375, 307, 253, 208, 172, 142, 118, 98, 82, 69, 58].
\end{itemize}

%% file: refs.bib
@article{cover1973enumerative,
  author  = {Thomas M. Cover},
  title   = {Enumerative Source Encoding},
  journal = {IEEE Transactions on Information Theory},
  volume  = {19},
  number  = {1},
  pages   = {73--77},
  year    = {1973},
  doi     = {10.1109/TIT.1973.1054929}
}

@article{flajolet1985pcsa,
  author  = {Philippe Flajolet and G. Nigel Martin},
  title   = {Probabilistic Counting Algorithms for Data Base Applications},
  journal = {Journal of Computer and System Sciences},
  volume  = {31},
  number  = {2},
  pages   = {182--209},
  year    = {1985},
  doi     = {10.1016/0022-0000(85)90041-8}
}

@article{hammer1995horn,
  author  = {Peter L. Hammer and Alexander Kogan},
  title   = {Quasi-Acyclic Propositional {Horn} Knowledge Bases: Optimal Compression},
  journal = {IEEE Transactions on Knowledge and Data Engineering},
  volume  = {7},
  number  = {5},
  pages   = {751--762},
  year    = {1995},
  doi     = {10.1109/69.469822}
}

@inproceedings{durand2003loglog,
  author    = {Marianne Durand and Philippe Flajolet},
  title     = {Loglog Counting of Large Cardinalities},
  booktitle = {Proceedings of the 11th Annual European Symposium on Algorithms (ESA)},
  pages     = {605--617},
  publisher = {Springer},
  address   = {Berlin, Heidelberg},
  year      = {2003},
  doi       = {10.1007/978-3-540-39658-1_55}
}

@inproceedings{estan2003bitmap,
  author    = {Cristian Estan and George Varghese and Michael E. Fisk},
  title     = {Bitmap Algorithms for Counting Active Flows on High Speed Links},
  booktitle = {Proceedings of the 3rd ACM SIGCOMM Conference on Internet Measurement (IMC)},
  pages     = {153--166},
  publisher = {ACM},
  address   = {New York, NY, USA},
  year      = {2003},
  doi       = {10.1145/948224.948225}
}

@inproceedings{scheuermann2007compression,
  author    = {Bj{\"o}rn Scheuermann and Martin Mauve},
  title     = {Near-Optimal Compression of Probabilistic Counting Sketches for Networking Applications},
  booktitle = {Proceedings of the 4th ACM International Workshop on Foundations of Mobile Computing (DIALM-POMC)},
  pages     = {1--7},
  publisher = {ACM},
  address   = {New York, NY, USA},
  year      = {2007}
}

@inproceedings{flajolet2007hll,
  author    = {Philippe Flajolet and {\'E}ric Fusy and Olivier Gandouet and Fr{\'e}d{\'e}ric Meunier},
  title     = {{HyperLogLog}: The Analysis of a Near-Optimal Cardinality Estimation Algorithm},
  booktitle = {Proceedings of the International Conference on Analysis of Algorithms (AOFA)},
  series    = {DMTCS Proceedings},
  volume    = {AH},
  pages     = {127--146},
  publisher = {DMTCS},
  address   = {Nancy, France},
  year      = {2007},
  doi       = {10.46298/dmtcs.3545}
}

@inproceedings{kane2010optimal,
  author    = {Daniel M. Kane and Jelani Nelson and David P. Woodruff},
  title     = {An Optimal Algorithm for the Distinct Elements Problem},
  booktitle = {Proceedings of the 29th ACM SIGMOD-SIGACT-SIGART Symposium on Principles of Database Systems (PODS)},
  pages     = {41--52},
  publisher = {ACM},
  address   = {New York, NY, USA},
  year      = {2010},
  doi       = {10.1145/1807085.1807094}
}

@inproceedings{heule2013hllpp,
  author    = {Stefan Heule and Marc Nunkesser and Alexander Hall},
  title     = {{HyperLogLog} in Practice: Algorithmic Engineering of a State of the Art Cardinality Estimation Algorithm},
  booktitle = {Proceedings of the 16th International Conference on Extending Database Technology (EDBT)},
  pages     = {683--692},
  publisher = {ACM},
  address   = {New York, NY, USA},
  year      = {2013},
  doi       = {10.1145/2452376.2452456}
}

@inproceedings{cohen2014hip,
  author    = {Edith Cohen},
  title     = {All-Distances Sketches, Revisited: {HIP} Estimators for Massive Graphs Analysis},
  booktitle = {Proceedings of the 33rd ACM SIGMOD-SIGACT-SIGART Symposium on Principles of Database Systems (PODS)},
  pages     = {88--99},
  publisher = {ACM},
  address   = {New York, NY, USA},
  year      = {2014},
  doi       = {10.1145/2594538.2594546}
}

@inproceedings{ting2014streamed,
  author    = {Daniel Ting},
  title     = {Streamed Approximate Counting of Distinct Elements: Beating Optimal Batch Methods},
  booktitle = {Proceedings of the 20th ACM SIGKDD International Conference on Knowledge Discovery and Data Mining (KDD)},
  pages     = {442--451},
  publisher = {ACM},
  address   = {New York, NY, USA},
  year      = {2014},
  doi       = {10.1145/2623330.2623669}
}

@misc{ertl2017hll,
  author        = {Otmar Ertl},
  title         = {New Cardinality Estimation Algorithms for {HyperLogLog} Sketches},
  eprint        = {1702.01284},
  archivePrefix = {arXiv},
  year          = {2017}
}

@misc{lang2017cpc,
  author        = {Kevin J. Lang},
  title         = {Back to the Future: An Even More Nearly Optimal Cardinality Estimation Algorithm},
  eprint        = {1708.06839},
  archivePrefix = {arXiv},
  year          = {2017}
}

@article{johansson2017arb,
  author  = {Fredrik Johansson},
  title   = {Arb: Efficient Arbitrary-Precision Midpoint-Radius Interval Arithmetic},
  journal = {IEEE Transactions on Computers},
  volume  = {66},
  number  = {8},
  pages   = {1281--1292},
  year    = {2017},
  doi     = {10.1109/TC.2017.2690633}
}

@misc{collet2021zstd,
  author       = {Yann Collet and Murray Kucherawy},
  title        = {Zstandard Compression and the `application/zstd' Media Type},
  howpublished = {IETF RFC 8878},
  year         = {2021},
  doi          = {10.17487/RFC8878}
}

@article{ertl2021setsketch,
  author  = {Otmar Ertl},
  title   = {{SetSketch}: Filling the Gap between {MinHash} and {HyperLogLog}},
  journal = {Proceedings of the VLDB Endowment},
  volume  = {14},
  number  = {11},
  pages   = {2244--2257},
  year    = {2021},
  doi     = {10.14778/3476249.3476276}
}

@misc{ohayon2021ehll,
  author        = {Tal Ohayon},
  title         = {{ExtendedHyperLogLog}: Analysis of a New Cardinality Estimator},
  eprint        = {2106.06525},
  archivePrefix = {arXiv},
  year          = {2021}
}

@inproceedings{pettie2021fish,
  author    = {Seth Pettie and Dingyu Wang},
  title     = {Information Theoretic Limits of Cardinality Estimation: {Fisher} Meets {Shannon}},
  booktitle = {Proceedings of the 53rd Annual ACM SIGACT Symposium on Theory of Computing (STOC)},
  pages     = {556--569},
  publisher = {ACM},
  address   = {New York, NY, USA},
  year      = {2021},
  doi       = {10.1145/3406325.3451032},
  note      = {Extended version: arXiv:2007.08051v3}
}

@inproceedings{pettie2021nonmergeable,
  author    = {Seth Pettie and Dingyu Wang and Longhui Yin},
  title     = {Non-Mergeable Sketching for Cardinality Estimation},
  booktitle = {Proceedings of the 48th International Colloquium on Automata, Languages, and Programming (ICALP)},
  series    = {LIPIcs},
  volume    = {198},
  pages     = {104:1--104:20},
  publisher = {Schloss Dagstuhl -- Leibniz-Zentrum f{\"u}r Informatik},
  address   = {Dagstuhl, Germany},
  year      = {2021},
  doi       = {10.4230/LIPIcs.ICALP.2021.104}
}

@inproceedings{karppa2022hlll,
  author    = {Matti Karppa and Rasmus Pagh},
  title     = {{HyperLogLogLog}: Cardinality Estimation With One Log More},
  booktitle = {Proceedings of the 28th ACM SIGKDD Conference on Knowledge Discovery and Data Mining (KDD)},
  pages     = {753--761},
  publisher = {ACM},
  address   = {New York, NY, USA},
  year      = {2022},
  doi       = {10.1145/3534678.3539246}
}

@inproceedings{wang2023gra,
  author    = {Dingyu Wang and Seth Pettie},
  title     = {Better Cardinality Estimators for {HyperLogLog}, {PCSA}, and Beyond},
  booktitle = {Proceedings of the 42nd ACM SIGMOD-SIGACT-SIGAI Symposium on Principles of Database Systems (PODS)},
  pages     = {317--327},
  publisher = {ACM},
  address   = {New York, NY, USA},
  year      = {2023},
  doi       = {10.1145/3584372.3588680}
}

@inproceedings{karayel2023optimal,
  author    = {Emin Karayel},
  title     = {An Embarrassingly Parallel Optimal-Space Cardinality Estimation Algorithm},
  booktitle = {Approximation, Randomization, and Combinatorial Optimization. Algorithms and Techniques (APPROX/RANDOM)},
  series    = {LIPIcs},
  volume    = {275},
  pages     = {35:1--35:22},
  publisher = {Schloss Dagstuhl -- Leibniz-Zentrum f{\"u}r Informatik},
  address   = {Dagstuhl, Germany},
  year      = {2023},
  doi       = {10.4230/LIPIcs.APPROX/RANDOM.2023.35}
}

@inproceedings{du2023spikesketch,
  author    = {Yang Du and He Huang and Yu-E Sun and Kejian Li and Boyu Zhang and Guoju Gao},
  title     = {A Better Cardinality Estimator with Fewer Bits, Constant Update Time, and Mergeability},
  booktitle = {Proceedings of the IEEE Conference on Computer Communications (INFOCOM)},
  pages     = {1--10},
  publisher = {IEEE},
  address   = {Piscataway, NJ, USA},
  year      = {2023},
  doi       = {10.1109/INFOCOM53939.2023.10229088}
}

@article{ertl2024ull,
  author  = {Otmar Ertl},
  title   = {{UltraLogLog}: A Practical and More Space-Efficient Alternative to {HyperLogLog} for Approximate Distinct Counting},
  journal = {Proceedings of the VLDB Endowment},
  volume  = {17},
  number  = {7},
  pages   = {1655--1668},
  year    = {2024},
  doi     = {10.14778/3654621.3654632}
}

@inproceedings{ertl2025ell,
  author    = {Otmar Ertl},
  title     = {{ExaLogLog}: Space-Efficient and Practical Approximate Distinct Counting up to the Exa-Scale},
  booktitle = {Proceedings of the 28th International Conference on Extending Database Technology (EDBT)},
  pages     = {829--841},
  publisher = {OpenProceedings.org},
  address   = {Konstanz, Germany},
  year      = {2025},
  doi       = {10.48786/edbt.2025.67}
}

@article{janson2025hypertwobits,
  author  = {Svante Janson and J{\'e}r{\'e}mie Lumbroso and Robert Sedgewick},
  title   = {Bit-Array-Based Alternatives to {HyperLogLog}},
  journal = {Theoretical Computer Science},
  volume  = {1054},
  pages   = {115450},
  year    = {2025},
  doi     = {10.1016/j.tcs.2025.115450}
}

@phdthesis{wang2025thesis,
  author = {Dingyu Wang},
  title  = {Streaming Sketching: Mathematical Theory and Practical Algorithms},
  school = {University of Michigan},
  year   = {2025}
}

@misc{karppa2026hbs,
  author        = {Matti Karppa},
  title         = {{Huffman-Bucket Sketch}: A Simple $O(m)$ Algorithm for Cardinality Estimation},
  eprint        = {2603.10930},
  archivePrefix = {arXiv},
  year          = {2026}
}

@misc{bushnell2026dll,
  author        = {Brian Bushnell},
  title         = {{DynamicLogLog}: Faster, Smaller, and More Accurate Cardinality Estimation},
  eprint        = {2603.27405},
  archivePrefix = {arXiv},
  year          = {2026}
}

@misc{bushnell2026avll,
  author        = {Brian Bushnell},
  title         = {Arithmetic Variable {LogLog}: Advancing the Memory-Variance Frontier},
  eprint        = {2608.12575},
  archivePrefix = {arXiv},
  year          = {2026}
}

@misc{datasketches,
  author       = {{Apache Software Foundation}},
  title        = {Apache {DataSketches} C++ Library (v5.2.0)},
  howpublished = {\url{https://github.com/apache/datasketches-cpp}},
  year         = {2025}
}

@article{xiao2020tailcut,
  author  = {Qingjun Xiao and Shigang Chen and You Zhou and Junzhou Luo},
  title   = {Estimating Cardinality for Arbitrarily Large Data Stream With Improved Memory Efficiency},
  journal = {IEEE/ACM Transactions on Networking},
  volume  = {28},
  number  = {2},
  pages   = {433--446},
  year    = {2020},
  doi     = {10.1109/TNET.2020.2970860}
}

@inproceedings{blasiok2018optimal,
  author    = {Jaros{\l}aw B{\l}asiok},
  title     = {Optimal Streaming and Tracking Distinct Elements with High Probability},
  booktitle = {Proceedings of the Twenty-Ninth Annual ACM-SIAM Symposium on Discrete Algorithms (SODA)},
  pages     = {2432--2448},
  publisher = {SIAM},
  year      = {2018},
  doi       = {10.1137/1.9781611975031.156}
}

@misc{duda2013ans,
  author        = {Jarek Duda},
  title         = {Asymmetric Numeral Systems: Entropy Coding Combining Speed of {Huffman} Coding with Compression Rate of Arithmetic Coding},
  year          = {2013},
  eprint        = {1311.2540},
  archivePrefix = {arXiv},
  primaryClass  = {cs.IT}
}

@article{elias1974efficient,
  author  = {Peter Elias},
  title   = {Efficient Storage and Retrieval by Content and Address of Static Files},
  journal = {Journal of the ACM},
  volume  = {21},
  number  = {2},
  pages   = {246--260},
  year    = {1974},
  doi     = {10.1145/321812.321820}
}
